\documentclass[superscriptaddress,showpacs,showkeys,twocolumn,aps,prb]{revtex4-1}
\usepackage{graphicx}% Include figure files
\usepackage{dcolumn}% Align table columns on decimal point
\usepackage{bm}% bold math
\usepackage{mathbbol}
\usepackage{bbm}
\usepackage{amsmath,amssymb,amsthm}
\usepackage{amsfonts}
\usepackage{hyperref}
\hypersetup{
    colorlinks=true,       % false: boxed links; true: colored links
    linkcolor=cyan,          % color of internal links
    citecolor=magenta,        % color of links to bibliography
    filecolor=magenta,      % color of file links
    urlcolor=cyan,           % color of external links
    runcolor=cyan
}
\usepackage[svgnames]{xcolor}

\usepackage{times}
\usepackage{comment}
\usepackage{graphicx}
\usepackage{bm}
\usepackage{color}
\usepackage{multi row} 
\usepackage{ulem}

\begin{document}

\title{Localization transition induced by programmable disorder}
\author{Jaime L. C. da C. Filho}
\email{jaime.filho@ifpa.edu.br }
\affiliation{Instituto Federal do Par\'{a} - Campus Castanhal, BR 316 Km 61, Saudade II, 68740-970, Castanhal, PA, Brazil}
\author{Zoe Gonzalez Izquierdo}
\email{zgonzalez@usra.edu}
\affiliation{QuAIL, NASA Ames Research Center, Moffett Field, California 94035, USA}
\affiliation{USRA Research Institute for Advanced Computer Science, Mountain View, California 94043, USA}
\author{Andreia Saguia}
\email{asaguia@id.uff.br}
\affiliation{Instituto de F\'{i}sica, Universidade Federal Fluminense, Av. Gal. Milton Tavares de Souza s/n, Gragoat\'{a}, 24210-346 Niter\'{o}i, Rio de Janeiro, Brazil}
\author{Tameem Albash}
\email{talbash@unm.edu}
\affiliation{Department of Electrical and Computer Engineering,
University of New Mexico, Albuquerque, New Mexico 87131, USA}
\affiliation{Department of Physics and Astronomy and Center for Quantum Information and Control,
CQuIC, University of New Mexico, Albuquerque, New Mexico 87131, USA}
\author{Itay Hen}
\email{itayhen@isi.edu}
\affiliation{Department of Physics and Astronomy, and Center for Quantum Information Science \& Technology,
University of Southern California, Los Angeles, California 90089, USA}
\affiliation{Information Sciences Institute, University of Southern California, Marina del Rey, CA 90292, USA}
\author{Marcelo S. Sarandy}
\email{msarandy@id.uff.br}
\affiliation{Instituto de F\'{i}sica, Universidade Federal Fluminense, Av. Gal. Milton Tavares de Souza s/n, Gragoat\'{a}, 24210-346 Niter\'{o}i, Rio de Janeiro, Brazil}

\begin{abstract}
\noindent We investigate the occurrence of many-body localization (MBL) on a spin-1/2 transverse-field Ising model defined on a Chimera connectivity graph with  
random exchange interactions and longitudinal fields. We observe a transition from an ergodic 
phase to a non-thermal phase for individual energy eigenstates induced by a critical disorder strength for the Ising parameters. Our result follows from the analysis of both the mean half-system block entanglement and the energy level statistics. 
We identify the critical point associated with this transition using the maximum variance of the block entanglement over the disorder ensemble 
as a function of the disorder strength. The calculated energy density phase diagram shows the existence of a mobility edge in the energy spectrum. 
In terms of the energy level statistics, the system changes from the Gaussian orthogonal ensemble for weak disorder to a Poisson distribution limit for strong randomness, 
which implies localization behavior. 
We then realize the time-independent disordered Ising Hamiltonian experimentally using a reverse annealing quench-pause-quench protocol on a D-Wave 2000Q programmable quantum annealer. 
We characterize the transition from the thermal to the localized phase through magnetization measurements at 
the end of the annealing dynamics, and the results are compatible with our theoretical prediction for the critical point. However, the same behavior can be reproduced using a classical spin-vector Monte Carlo simulation, which suggests 
that genuine quantum signatures of the phase transition remain out of reach using this experimental platform and protocol.
\end{abstract}

\maketitle

%%%%%%%%%%%%%%
\section{Introduction}
%%%%%%%%%%%%%%
Many-body localization (MBL) is a remarkable quantum phenomenon induced by random coupling disorder. It has long been established 
that non-interacting quantum systems may spatially localize in the presence of uncorrelated~\cite{Anderson1958} or quasiperiodic~\cite{Aubry1980} 
onsite disorder. This phenomenon, known as {\it Anderson localization}, drives the system to an insulating phase. Analogously, MBL refers to 
localization in Hilbert space of interacting systems in the presence of disorder~\cite{Abanin:19}. 
The appearance of MBL behavior can be understood 
as a dynamical phase transition~\cite{DQPT,DQPT2}, where the energy eigenstates individually undergo a sharp change as the disorder strength is 
varied~\footnote{The concept of dynamical quantum phase transitions can be generically defined through the non-analytical behavior in time of the Loschmidt echo~\cite{DQPT,DQPT2}. Concerning the MBL transition, the term `dynamical' 
is often associated with the collective behavior shown by the set of energy eigenstates in a non-equilibrium evolution.}. 
This is similar to a conventional quantum phase transition, where the ground state changes significantly as the control parameter is varied across the critical point. 
The interplay between interaction and disorder had already been theoretically predicted~\cite{Anderson1958,Fleishman1980,Giamarchi1988} 
with perturbative methods confirming the existence of localized states at low temperatures in later sudies~\cite{Gornyi2005,Basko2006}. 
Full theoretical~\cite{Santos2005loc,Oganesyan2007,Znidaric2008,Pal2010,Bardarson:12,Bauer:13,
Kjall:14,Huse2014,Torres2015,Luitz2015,Serbyn2015,Imbrie2016a,Bera2016,Gogolin2016,Khemani2016,LuitzARXIV,Tomasi2017,Wahl:19} 
and experimental~\cite{Smith2015,Schreiber2015,Kondov2015,WeiARXIV,Xu:18} studies have since confirmed the existence of MBL phases 
through different physical architectures. 

Concurrently, quantum annealing (QA) optimizers~\cite{Kadowaki:98,Farhi:01} consisting of an array of
superconducting quantum interference device (SQUID) quantum bits (qubits)~\cite{Johnson:11} have increasingly become an experimental platform 
to simulate properties of disordered condensed matter quantum systems~\cite{King:18,Harris:2018,Nishimura:20,Kairys:20,King:19,King:21}. 
QA exploits the gradual decrease of quantum-mechanical fluctuations to drive a quantum system to a target state that encodes 
the global minimum of a programmable objective function. 
Proposals to examine an MBL phase in quantum annealers have been introduced in recent years. 
This includes an order parameter for the MBL phase~\cite{Mozgunov:17} and 
tests of the MBL behavior for the implementation of the graph coloring algorithm~\cite{Kudo:20}. 
Here, we will explore a different direction, using a D-Wave 2000Q (DW2kQ) quantum annealer as a platform for the investigation of disorder-induced critical behavior through a general QA process.  The programmability of the device suggests that it might be amenable to study disorder-induced transitions as the local fields and interactions can be individually programmed, providing a setup where onsite disorder distributions can be realized.  In this work, we will investigate to what extent we can characterize the onset of a non-thermal localized phase on such devices.

We will focus our study on the Chimera connectivity graph~\cite{Choi:11} of the DW2kQ, whose topology will be shown to induce a non-trivial phase diagram driven by random disorder 
either in the Ising interactions or in the longitudinal local fields. This phase diagram exhibits a phase transition that separates an ergodic phase, in which the eigenstate
thermalization hypothesis (ETH) is obeyed~\cite{Deutsch:91,Srednicki:94,Rigol:08}, from a non-thermal phase, where 
memory of the initial configuration indicates localization behavior in Hilbert space. 
We will first show that block entanglement --- the von Neumann entropy of the reduced density matrix of a subsystem --- can 
help identify the phase transition through 
a change in the mean half-system block entropy for individual energy eigenstates. Specifically, we will show that an approximately disorder-independent 
block entropy that is typical of the 
ergodic phase undergoes a strong decrease as the system is driven to the non-thermal phase.
The reason for this is that localized states are concentrated 
in small regions of Hilbert space, which in turn results in small block entanglement as a function of disorder~\cite{Bauer:13,Kjall:14,Luitz2015,Bera2016}. 
We will characterize the critical point using the maximum 
variance of the mean block entanglement over the disorder ensemble as a function of the disorder strength~\cite{Kjall:14}. 
This quantity signals that, close to the critical point, block entanglement prominently fluctuates, since the system is at the 
border of a scaling behavior change.  
For an analysis of the critical 
point throughout the energy spectrum, we will also explicitly map out the energy density phase diagram. This procedure shows the existence of a many-body mobility edge~\cite{Basko2006,Luitz2015,Mott:75}, 
which implies the change of the properties of individual eigenstates as the energy density varies as a function of disorder. The localization behavior is also explored through the energy level statistics. 
Specifically, we will show that the distribution of mean energy gap ratios changes from the Gaussian orthogonal ensemble (GOE) for weak disorder to a Poisson distribution limit for strong randomness, 
implying a localized behavior.

Our experimental study of the phase transition on a physical quantum annealer is done by realizing time-independent disordered Ising models, 
which requires full control over the disorder distribution ensemble. Such control is achieved by exploiting two features of the DW2kQ: i) reverse annealing~\cite{Passarelli2020}, 
which is used to start the evolution with the system in an eigenstate of a classical Ising model, instead of the ground state of the transverse field Hamiltonian as in the usual QA approach, 
and ii) annealing pause, which we use to stop the evolution at a suitable dimensionless pause point $s_p \in (0,1)$, which sets the 
disorder strength. {Unlike the standard annealing protocol, the reverse annealing protocol allows us to start in an arbitrary classical Ising state and hence control the magnetization of the initial state.} 
Then, the dynamics of a time-independent (paused) disordered transverse-field Ising 
model takes place, {with $s_p$ controlling the amount of disorder in the paused Hamiltonian}. 
After the end of the pause, the asymptotic magnetization is then measured after a quench to the Ising Hamiltonian~\cite{King:18,Harris:2018,King:19,Nishimura:20}. 
The QA dynamics occurs under decoherence, so that the system is not expected to perfectly evolve as an eigenstate of the instantaneous Hamiltonian, 
as would be the case in a long-time adiabatic evolution occurring in a closed system. Thus, environmental noise and fast dynamics will spread state population throughout the energy spectrum. The localized phase is expected to exhibit a 
memory effect, `remembering' the initial value of the local magnetization. We will observe this memory effect at a disorder strength compatible with the theoretical
critical point. We will  also show that these features can be reproduced using a purely classical model of the system based on spin-vector Monte Carlo (SVMC), which then suggests 
that genuine quantum signatures for the localization transition remain inaccessible using this experimental platform and protocol.

%%%%%%%%%%%%%%%%%%%%%%%%%%%%%%
\section{Disordered spin-1/2 transverse-field Ising model on a Chimera connectivity graph}
\label{sec:tfim}
%%%%%%%%%%%%%%%%%%%%%%%%%%%%%%

\begin{figure}[htb]
\includegraphics[scale=0.3]{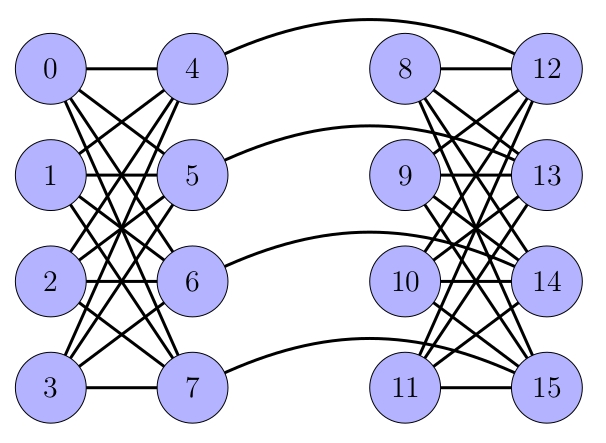}
\caption{Two horizontally-connected Chimera cells. 
Each unit cell consists of $N=8$ spins with bipartite intra-connectivity. The adjacent cells are inter-connected through one of the disjoint parts of the unit cell. 
Only horizontal inter-cell couplings are shown in the figure.}
\label{f1}
\end{figure}

We first examine the onset of a localized phase in transverse-field Ising models defined on a Chimera connectivity graph. The Hamiltonian is given by
\begin{equation}
H_{\textrm{TFI}} = H_{\textrm{TF}} + H_{\textrm{I}}, 
\label{htfi}
\end{equation}
where
\begin{eqnarray}
H_{\textrm{TF}}=  - \Delta \sum_{k} \sigma_{k}^{x} \,\, \textrm{and} \hspace{0.2cm} H_{\textrm{I}}= \sum_{\langle ij\rangle}J_{ij} \sigma_{i}^{z}\sigma_{j}^{z} + \sum_{i} h_i \sigma_i^z  ,\nonumber \\
\label{htfi2}
\end{eqnarray}
with $\sigma^{\alpha}_{k}$ denoting the Pauli operators in the direction $\alpha \in \{x,z\}$ on site $k$.  The transverse field strength $\Delta$ is fixed and both the 
exchange interactions $J_{ij}$  and longitudinal fields $h_i$ are random numbers taken from a uniform distribution in the interval $\left[-J,+J\right]$, with the indices 
$\langle ij\rangle$ running over the Chimera connectivity. The Chimera architecture in the D-Wave device comprises an array of unit cells, with each cell consisting of $N=8$ spins with a bipartite 
intra-connectivity, as shown in Fig.~\ref{f1}. Each cell is a complete bipartite $K_{4,4}$ graph, wherein each qubit is connected to four others within its cell and to two 
more in adjacent cells. {In Fig.~\ref{f1}, we have only displayed the horizontal (and not the vertical) inter-cell connections in the Chimera architecture 
(for a complete view of the DW2KQ hardware graph, see, e.g., Ref.~\cite{Albash:PRX2018}). }

%%%%%%%%%%%%%%%%%%%
\subsection{Entanglement signature of localization transition}
%%%%%%%%%%%%%%%%%%%
 
The localization transition that we are interested in has a dynamical nature, which can be probed by studying the quantum correlations in the individual many-body energy eigenstates~\cite{Pal2010,Kjall:14}. 
We do this by studying the half-system block entanglement. For low disorder (small $J / \Delta$), the block entanglement is approximately disorder-independent, which is consistent with an ergodic 
behavior for the system. On the other hand, as we increase disorder, the system is driven to the localized phase, 
with an observed strong decrease of block entanglement. {The disorder strength $J/\Delta$ is non-vanishing, ensuring nonintegrability.}
We  measure block entanglement by the von Neumann entropy 
of the half-system reduced density operator. For a half-unit Chimera cell, 
an up-down bipartition corresponds to the subsets of qubits 
$A = \{0,1,4,5\}$ and $B = \{2,3,6,7\}$, 
as labeled in Fig.~\ref{f1}. Then, given an energy eigenstate $|\psi\rangle$ and the 
bipartition of the system into two halves $A$ and $B$, entanglement between $A$ and $B$ is measured 
by the von Neumann entropy $S_E$ of the
reduced density matrix of either block, i.e., 
\begin{equation}
S_E =-\text{Tr} \left( \rho_A \log_2 \rho_A \right) = -\text{Tr}
\left( \rho_B \log_2 \rho_B \right),  \label{vonNeumann}
\end{equation}
where $\rho_A=\text{Tr}_B \rho$ and $\rho_B = \text{Tr}_A \rho$ denote the
reduced density matrices of blocks $A$ and $B$, respectively, with $\rho=|\psi\rangle\langle \psi|$. 

\begin{figure}[htb]
\includegraphics[scale=0.3]{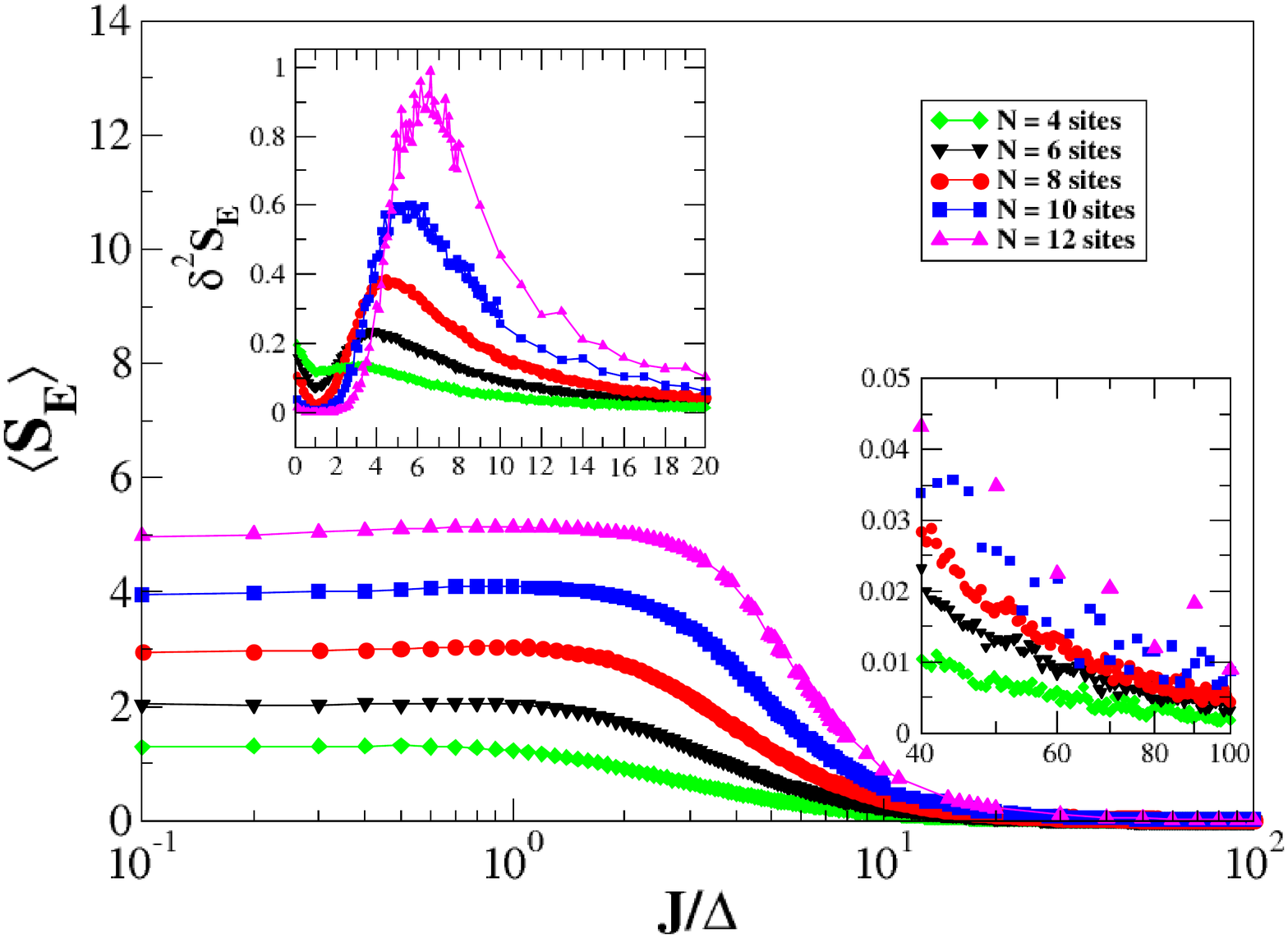}
\caption{Main panel: Mean half-system block entanglement for the eigenstate in the middle of the spectrum 
as a function of the disorder strength $J/\Delta$ for system sizes up to $N=12$ spins. 
Left inset: Variance of the mean block entanglement for disorder ensembles as a function of $J/\Delta$. Right inset: Block entanglement is zoomed in on the right-hand curve tail.}
\label{f2}
\end{figure}

In order to investigate entanglement, We  exactly diagonalize the Hamiltonian matrix for sizes initially up to a Chimera unit cell ($N=8$ spins) so that we can compare with  
experiments performed on a DW2kQ. In addition, we also extend the Chimera cell to $N=10$ ($K_{5,5}$ graph) and $N=12$  ($K_{6,6}$ graph) in the theoretical 
analysis as a further evidence for the critical behavior of the system. 
Our analysis is carried out for a single 
eigenstate in the middle of the energy spectrum, which is expected to be the hardest to localize. 
We perform averages over $5\times10^3$ disorder configurations {for $N=4$, $N=6$, and $N=8$ spins}. For $N=10$ and $N=12$ spins, we consider $1\times10^3$ and 
$5\times10^2$ disorder configurations, respectively. We then 
evaluate the average von Neumann entropy $\langle S_E \rangle$ for each cell size. 
The results are shown in Fig.~\ref{f2}.

For the region of low disorder $J / \Delta$, the mean entanglement $\langle S_E \rangle$ is approximately independent of the disorder strength. 
On the other hand, $\langle S_E \rangle$ changes its behavior for large disorder, showing a strong decrease typical of a localized phase. In the right inset, 
we zoom in on the curve tail. We can see that $\langle S_E \rangle$ still increases with the size $N$ of the system for large disorder. 
Notice that, with our partition for $A$ and $B$, there is no clear distinction between bulk or boundary since all qubits in $A$ interact with all qubits in $B$. 
Here, we will characterize the phase transition in terms of the disorder strength. 
In the left inset, we plot the variance of 
$S_E$ as a function of $J / \Delta$, which is defined by $\delta^2 S_E = \langle S_E^2 \rangle - \langle S_E \rangle^2$. 
{For a given system size and disorder strength, the exact value of the entropy 
depends on the specific disorder realization. For a set of states obtained from an 
ensemble of disorder realizations, we will have both extended and localized states (for a review, see, e.g., Ref.~\cite{Alet:18}). 
Then, close to the critical disorder strength, large deviations with respect to the mean value are 
expected in the entanglement entropy pattern, leading to a divergence in its variance~\cite{Kjall:14}. 
By considering the scaling with the system size, we observe 
that the range of disorder strengths showing this mixing of extended and localized states 
narrows, tending to a peak at the critical point in the thermodynamic limit. }
For a finite system, the critical behavior is reflected by 
a maximum value for $\delta^2 S_E$. This maximum, 
which is the precursor of the critical point, occurs at $(J / \Delta)_c \approx 4.8$ for a system with $N=8$ spins, as shown in the Fig.~\ref{f2} inset.
This is a key size in our analysis, since it corresponds to the unit Chimera cell in the DW2kQ. 
Indeed, as we show in Sec.~\ref{sec:exp}, a dynamic manifestation of the critical point appears through local magnetization measurements on the DW2kQ. 
Moreover, it is shown in Appendix~\ref{LR_partition} that this maximum value can also be achieved for different partitions of the Chimera cell. 

%%%%%%%%%%%%%%%%%%%%%%%%%%%%%%
\subsection{Energy density phase diagram and mobility edge}
\begin{figure}[!hbt]
\includegraphics[scale=0.31]{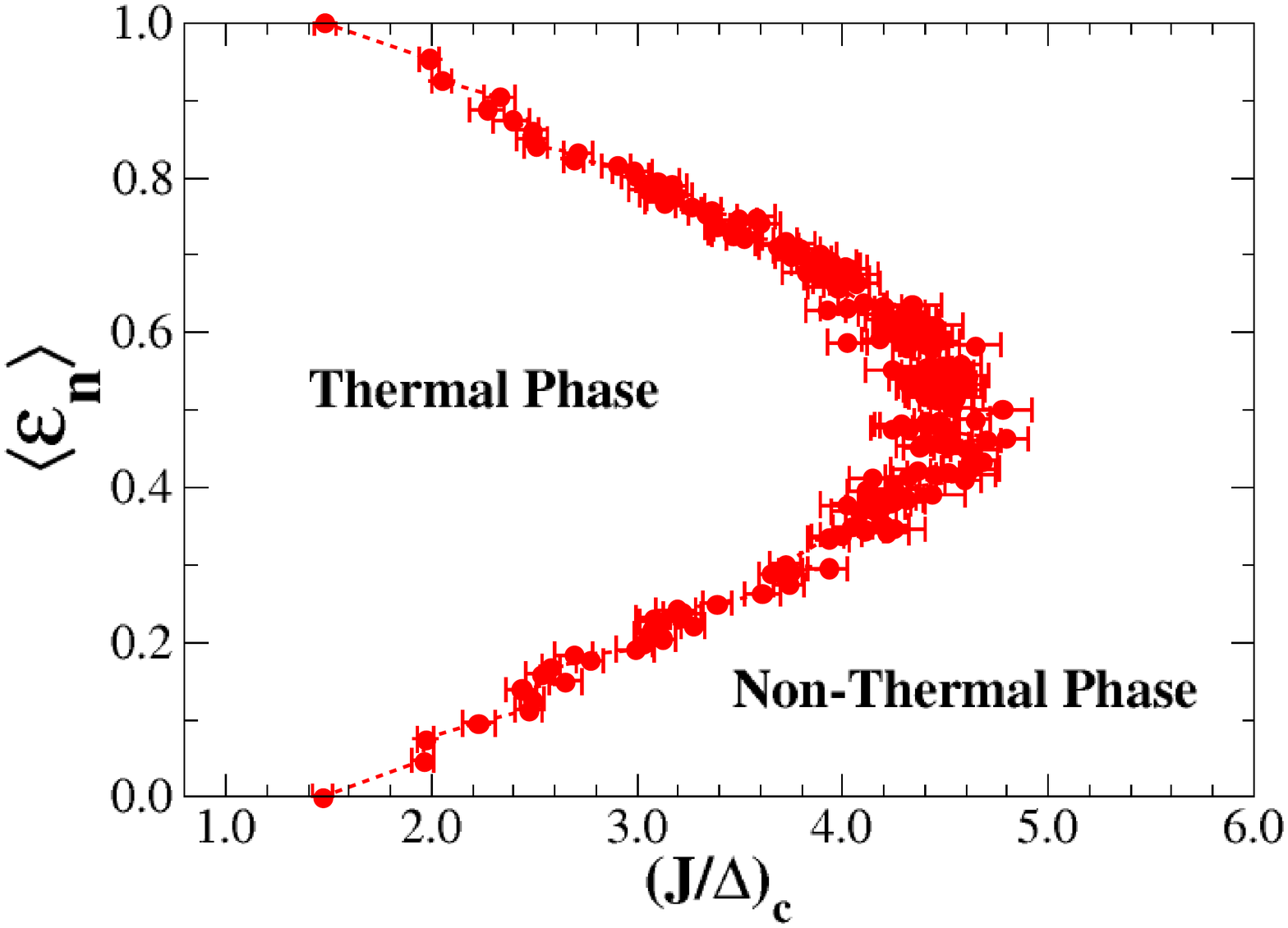}
\caption{{Energy density phase diagram for $N=8$ spins, with the critical mean energy density $\langle \varepsilon_n \rangle$ 
obtained from Eq.~(\ref{ene-me}) and averaged} as a function of the critical disorder $(J/\Delta)_c$.  
The ergodic and non-ergodic phases are separated by the presence of a mobility edge in the energy spectrum. Error bars are obtained by the standard deviation divided by the square root of the sample size.}
\label{f3}
\end{figure}

%%%%%%%%%%%%%%%%%%%%%%%%%%%%%%
The critical behavior of the variance of the entanglement ($\delta^2 S_E $) shown in the previous subsection for the middle-energy eigenstate generalizes throughout the energy spectrum. 
This can be conveniently analyzed through an energy density phase diagram, which we show exhibits a mobility edge when all the eigenstates 
are taken into account~\cite{Basko2006,Luitz2015,Mott:75}. For each disorder strength, we calculate the lowest and highest energies ($E_0$ and $E_{\textrm{max}}$ respectively) and then  
define the normalized energy density 
\begin{equation}
\varepsilon_n = \frac{E_{\textrm{max}}-E_n}{E_{\textrm{max}}-E_0}, 
\label{ene-me}
\end{equation}
where $E_n$ denotes the energy of an eigenlevel $n$. In particular, notice that we have $0 \le \varepsilon_n \le 1$, with $\varepsilon_n=0$ and $\varepsilon_n=1$ for the highest $E_{\textrm{max}}$ 
and lowest $E_0$ energy levels, respectively. 
{For each disorder realization, we obtain the eigenvalues and eigenvectors of the 
Hamiltonian, labelling them with an index $n$. Then, for each eigenstate, we compute 
the energy density $\varepsilon_n$ and the half-system block entropy.}
The energy density $\varepsilon_n$ is averaged over $5 \times 10^3$ realizations, 
which defines the mean energy density $\langle \varepsilon_n \rangle$. Doing so allows us to identify the 
critical disorder for each energy eigenstate using the maximum of $\delta^2 S_E$, 
as previously illustrated in Fig.~\ref{f2} for the middle-energy eigenstate. For the case of $N=8$ spins, the results for $\langle \varepsilon_n \rangle$ 
{as a function of the critical disorder strength $(J/\Delta)_c$} are shown in Fig.~\ref{f3}. 

We observe that the phase transition yields 
a boundary in the energy density diagram, with the appearance of a critical line as a function of the disorder strength. This is associated with the fact that the 
localization behavior arises as a dynamical transition undergone by each individual energy eigenstate. 
Notice also that the eigenstates in the middle of the spectrum 
are the hardest to localize. 
{Indeed, the convergence to the ETH behavior is easier to achieve in energy regions where we have a larger density of states, which typically occurs in the 
middle of the Hamiltonian spectrum. Since we have a large density of states, there is a higher probability to find eigenstates individually yielding similar expectation 
values to local observables. For the extreme parts of the spectrum, the density of states is usually small, making convergence to the ETH behavior slower (see, e.g., Ref.~\cite{Alet:18}).} 
The energy density phase diagram then implies the existence of a many-body mobility edge in the energy spectrum. 
This is characterized by a rearrangement in the entanglement properties of the eigenstates. 
The phase diagram conveys that for a fixed disorder and an energy density below a threshold value defined by the critical line, 
the quantum state is close to a product state of localized spins; as we cross it, the eigenstates become extended, with 
ETH taking place. Continuing to move vertically in the phase diagram, a new transition back 
to localization may occur as the critical line is crossed again.

%%%%%%%%%%%%%%%%%%%%%%%%%%%%%%
\subsection{Energy level statistics}
%%%%%%%%%%%%%%%%%%%%%%%%%%%%%%
In order to provide further evidence of localization in the non-thermal phase, we consider the energy spectrum statistics of finite-size samples.  
In the context of random matrix ensembles, it is expected that the energy level spacing exhibits a crossover from a GOE description 
in the diffusive regime to a Poisson distribution in the localization regime at strong randomness~\cite{Oganesyan2007,Pal2010}. 

\begin{figure}[!hbt]
\includegraphics[scale=0.31]{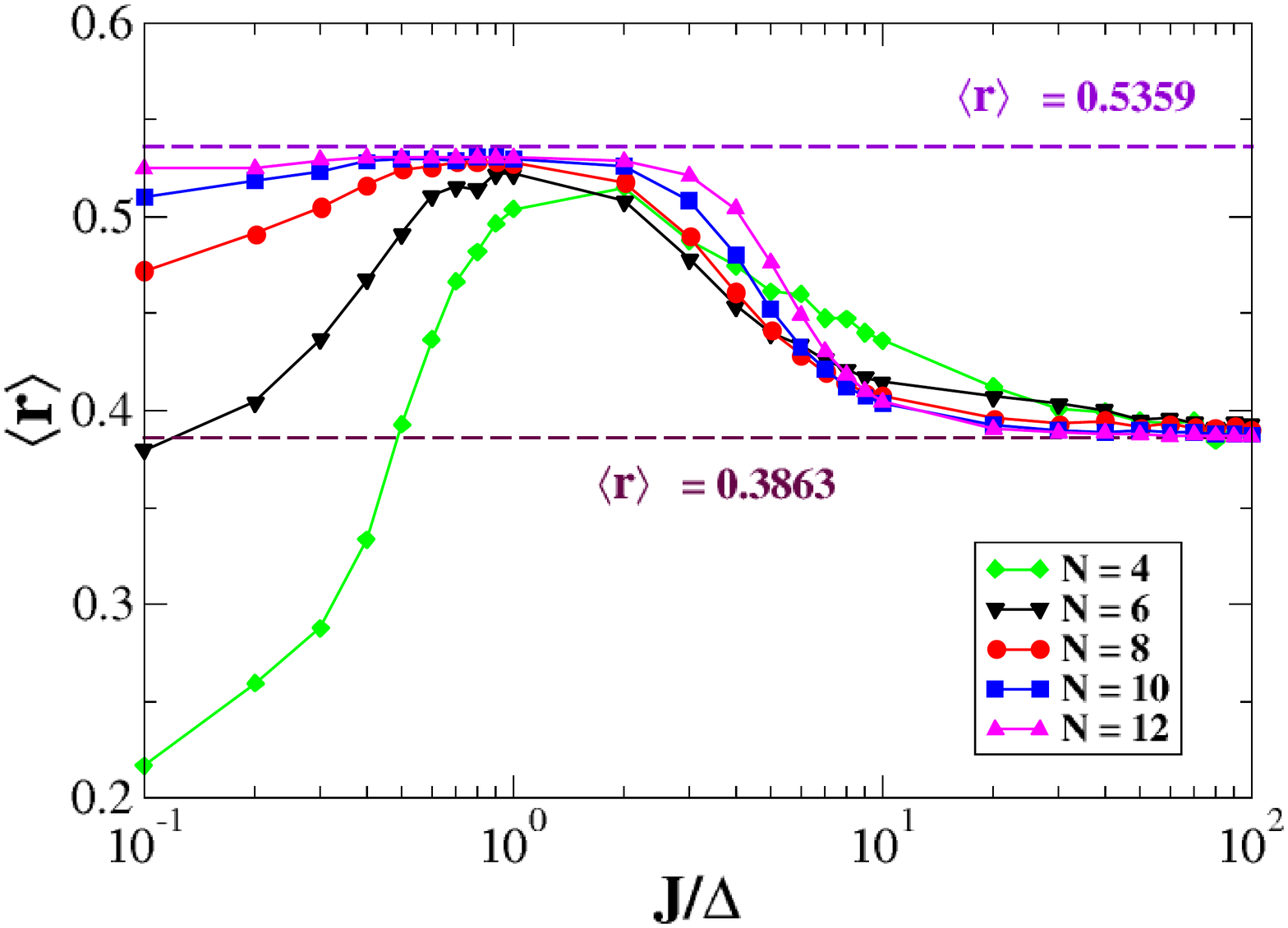}
\caption{Energy level statistics as a function of the disorder strength $J/\Delta$.  
The ergodic and non-thermal phases are characterized by different distribution limits in the regime of weak and strong disorder, respectively.}
\label{f40}
\end{figure}

For each disorder realization, we consider the energy level spacing $\delta_n = E_{n+1} - E_{n}$, where the energies $E_n$ are listed in ascending order. 
Given all the pairs $(\delta_n,\delta_{n+1})$, the ratio $r_n$ of adjacent energy gaps is defined as 
\begin{equation}
r_n = \frac{\min(\delta_n,\delta_{n+1})}{\max(\delta_n,\delta_{n+1})}.
\label{ration}
\end{equation}
We then average $r_n$ over all energy gaps and disorder samples, yielding $\langle r \rangle$. In the ergodic phase, the level statistics are expected to follow a GOE description, characterized by the distribution 
\begin{equation}
P_G = \frac{\pi}{2} \frac{\delta}{\langle \delta \rangle} \exp\left[- \frac{\pi \delta^2}{4\langle \delta \rangle^2}\right],
\end{equation}
where $\langle \delta \rangle$ is the mean spacing, which is obtained through the average of $\delta_n$ over $n$ (see, e.g., Ref.~\cite{Wang:21}). By assuming a GOE distribution, the Wigner-like surmise gives 
$\langle r \rangle = 4 - 2\sqrt{3} \approx 0.5359$~~\cite{Atas:13}. On the other hand, 
for a localized phase, the level statistics obeys a Poisson distribution
\begin{equation}
P_P =  \frac{1}{\langle \delta \rangle} \exp\left[- \frac{\delta}{\langle \delta \rangle}\right],
\end{equation}
with $\langle r \rangle = 2 \, \textrm{ln} 2 - 1 \approx 0.3863$~~\cite{Atas:13} (see also Refs.~\cite{Oganesyan2007,Pal2010,Wang:21}). 
This behavior is shown in Fig.~\ref{f40}, where {we numerically obtain $\langle r \rangle$ by averaging over the same number of disorder configurations as described in the previous subsections}. 
As we increase $N$, the average ratio $\langle r \rangle$ tends to a GOE in the weak $J/\Delta$ regime, 
whereas in the limit of strong order it approaches a Poisson distribution. This is an indication that the non-thermal phase obtained here for the TFI model on the Chimera graph indeed exhibits localized behavior.   

%%%%%%%%%%%%%%%%%%%%%%%%%%%%%%
\section{Experimental study of localization on a quantum annealer}
\label{sec:exp}
%%%%%%%%%%%%%%%%%%%%%%%%%%%%%%
In the previous section we studied analytically and numerically the localization transition for a Chimera cell, identifying the critical disorder value $(J / \Delta)_c \approx 4.8$ for energy eigenstates in the middle of the spectrum. 
We now proceed to experimentally observe a signature of this transition on a physical quantum annealer. While our theoretical results were calculated under the assumption of a closed, perfectly isolated system, it is unavoidable in any experimental setup to have some degree of coupling to the environment. Fortunately, characteristics of localization survive in open systems as long as the system-bath coupling is weaker than the other energy scales of the system~\cite{Huse2014}. Of course, even with weak system-bath coupling, for long enough timescales thermalization will eventually take place. But at intermediate timescales, between the relaxation time of the system and the thermalization with the environment, localization properties can be experimentally measured~\cite{Abanin:19}. 
The nonzero coupling to the surrounding environment also has the effect of broadening the localization transition into a crossover region, playing a similar role to nonzero temperature in quantum phase transitions~\cite{Luschen}. In this crossover region, the dynamics of ergodic and non-ergodic phases will smoothly interpolate. We therefore expect our experiments to reproduce a noisy version of the theoretical predictions.

For our experimental investigation, we use two DW2kQ devices so that we can analyze the effects of noise on the experimental results.
The two processors have the same architecture, but the fabrication process is improved between the two, allowing the newer one to attain a several-fold reduction in flux noise~\cite{dw_whitepaper_1}, one of the main sources of decoherence. This reduction in noise has been shown to increase tunneling rates, possibly leading to improved performance~\cite{dw_whitepaper_2}. We were performing our data collection on the noisier device when the newer, lower-noise one became available. We repeated our experiments on this improved version, and these newer results are the ones we present in this section. Although our main finding---the agreement between theory and experiment on the critical disorder at which memory effects emerge---remains consistent across both devices, we found certain differences in their behavior due to their disparate noise levels, so the results from the noisier device are reported in Appendix~\ref{high_noise}.

\begin{figure}[!hb]
\includegraphics[scale=0.4]{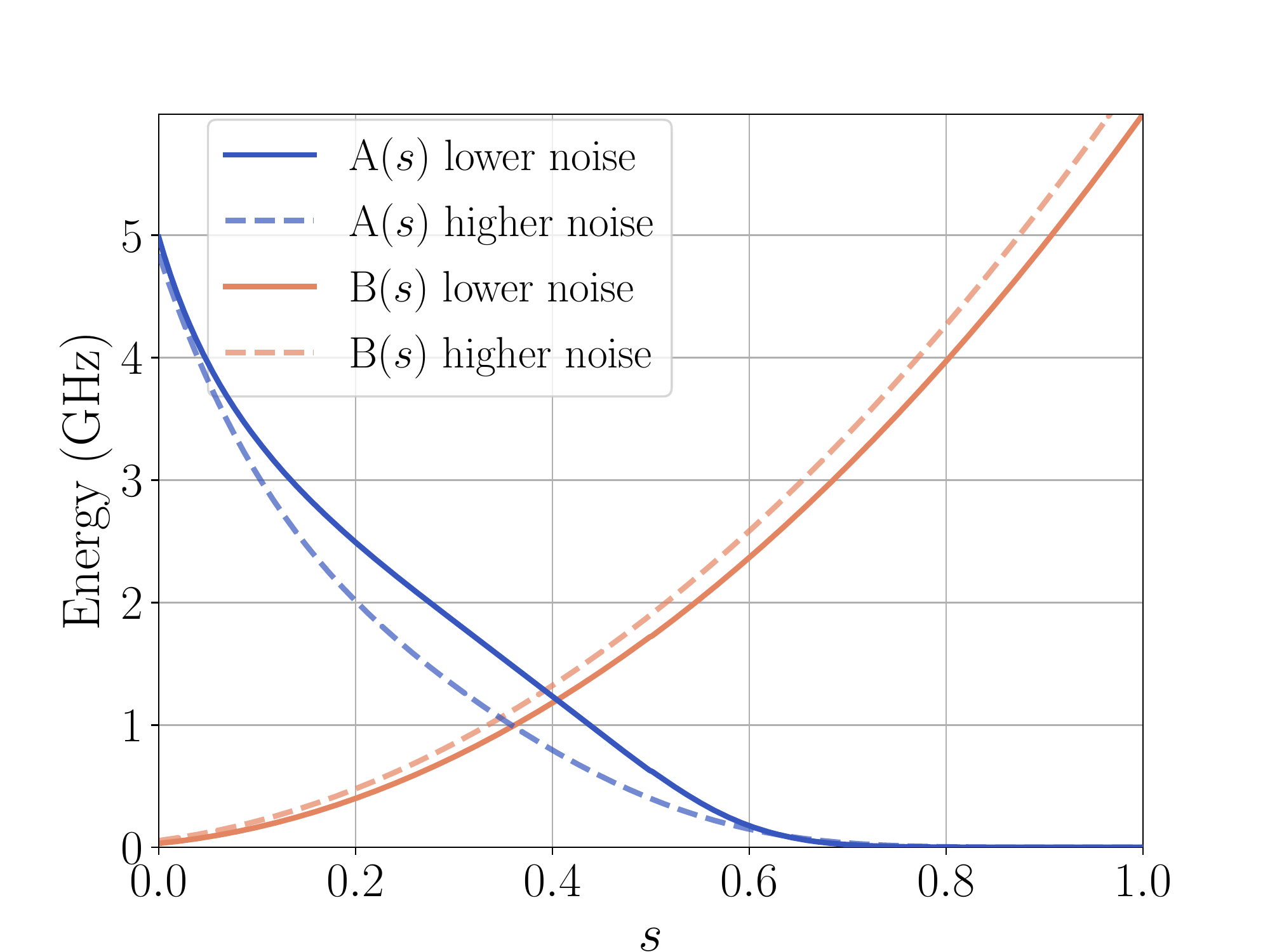}
\caption{The annealing schedules $\mathrm{A}(s)$ and $\mathrm{B}(s)$ as functions of the dimensionless time parameter $s$, 
in units of $h =1$. The schedules are plotted for both lower and higher noise processors.}
\label{f4}
\end{figure}

The DW2kQ implements a time-dependent Hamiltonian $H(s)$, where 
$s$ is a dimensionless time parameter, with $s \in [0,1]$. The Hamiltonian interpolates between the transverse-field contribution 
$H_{\textrm{TF}}$ and the classical Ising term $H_{\textrm{I}}$, reading
\begin{equation} \label{eqt:DWHamil}
H(s) = \mathrm{A}(s) H_{\textrm{TF}} +\mathrm{B}(s) H_{\textrm I}, 
\end{equation} 
where $H_{\textrm{TF}}$ and $H_{\textrm I}$ are provided by Eq.~(\ref{htfi2}), with $\mathrm{A}(s)$ and $\mathrm{B}(s)$ time-dependent functions determining the annealing 
schedule and fixed by the quantum hardware. The time dependence of $\mathrm{A}(s)$ and $\mathrm{B}(s)$ on $s$ is depicted in Fig.~\ref{f4}. 

We are, however, interested in a time-independent, disordered transverse-field Ising model, as described at the beginning of Sec.~\ref{sec:tfim}. To implement this, 
we use a reverse annealing protocol with a mid-anneal pause. Unlike the standard forward anneal, where the system starts at $s=0$ in the ground state of the 
transverse-field Hamiltonian $H_{\textrm{TF}}$ and then evolves towards $s=1$, here we prepare it in an eigenstate of the classical Ising Hamiltonian $H_{\textrm{I}}$ 
and evolve from the standard endpoint of $s=1$  to some pause location $s_p \in (0, 1)$, such that the ratio $\mathrm{B}(s_p)/\mathrm{A}(s_p)$ sets the disorder strength. 
At $s_p$, the system follows a pause-quench protocol, where the dynamics of a time-independent Hamiltonian first takes place for a fixed duration, and then the 
system evolves back to $s = 1$. The process is then concluded with a measurement in the computational basis, from which we can obtain a value for 
the magnetization $M_z  = \sum_i \sigma^z_i$.  
{We emphasize that the dimensionless pause point $s_p$ sets the disorder strength, and over the pause duration the system undergoes open-system dynamics with 
a time-independent Hamiltonian $H(s_p)$.}
We repeat the process $1000$ times (anneals) to obtain the statistics to estimate the average magnetization. 
Both the reverse and forward parts of the anneal are performed 
at the fastest rate allowed, namely $1 \mu s^{-1}$, and we choose a pause duration of $100 \mu s$. 

In our theoretical analysis (in Sec.~\ref{sec:tfim}), we used the half-system block entanglement to pinpoint the location of the critical point. While it is not possible to 
experimentally measure the same quantity on the DW2kQ, to find the critical disorder experimentally we can rely on the fact that, in localized phases, the memory of the 
initial state will persist in spite of the relaxation process. The above is true, for instance, for the mean (averaged over disorder) magnetization $\langle M_z \rangle$.  
On the DW2kQ, we obtain the mean magnetization as follows. For a fixed initial spin state (for example the all-up state), the state is  given by a set of distinct superpositions of the energy eigenstates of $H_{\textrm{TFI}}$.  (When we interpolate the Hamiltonian to the point $s_p$ where it is equal to $H_{\textrm{TFI}}$, the system is then described by some mixed state due the dynamics associated with the evolution.)
If $H_{\textrm{TFI}}$ is in the ergodic phase, the expectation value of the magnetization will evolve in time and ultimately reach  
the value predicted by the microcanonical ensemble. For a large enough sample of disorder realizations, we expect to have enough distinct initial superpositions of the eigenstates of $H_{\textrm{TFI}}$
such that all the computational basis states equally contribute to the magnetization when we average over the disorder. This then implies a vanishing mean magnetization. This picture follows even in the 
presence of very fast dephasing rates, where we do not have coherence between the energy eigenstates.  
Notice that the ETH leads to an effective description of an isolated system with mean energy $\langle E \rangle$ in a microcanonical ensemble. 
The microcanonical density matrix $\rho_{mc}$ assigns equal probability to every micro-state whose energy falls within a range centered at $\langle E \rangle$, with 
$\rho_{mc}$ then proportional to the identity. This implies the vanishing of $\langle M_z \rangle$ independently of the initial state for long time dynamics, since 
$\langle M_z \rangle = \textrm{Tr} \left( \rho_{mc} \, S^z \right) = 0$, with $S^z = \sum_i \sigma^z_i$. 
On the other hand, in the localized phase, we have distinct magnetization patterns depending on the initial state, with the
memory of this initial state increased as the disorder strength gets larger.  
In our experiment, We  characterize this picture by first running the anneal process for single Chimera-cell instances and 
then checking its validity for double-cell instances. We  use initial states with magnetization $\pm N$ (i.e., all spins pointing up in the $z$-direction or all spins pointing down) 
 to observe the memory effects characteristic of localization, while we use initial states with magnetization zero (equal number of up and down spins) to serve as a control, 
since the final magnetization is expected to be zero in that case regardless of whether the system is in an ergodic or localized phase.

\begin{figure}[!hb]
\centering
\includegraphics[scale=0.41]{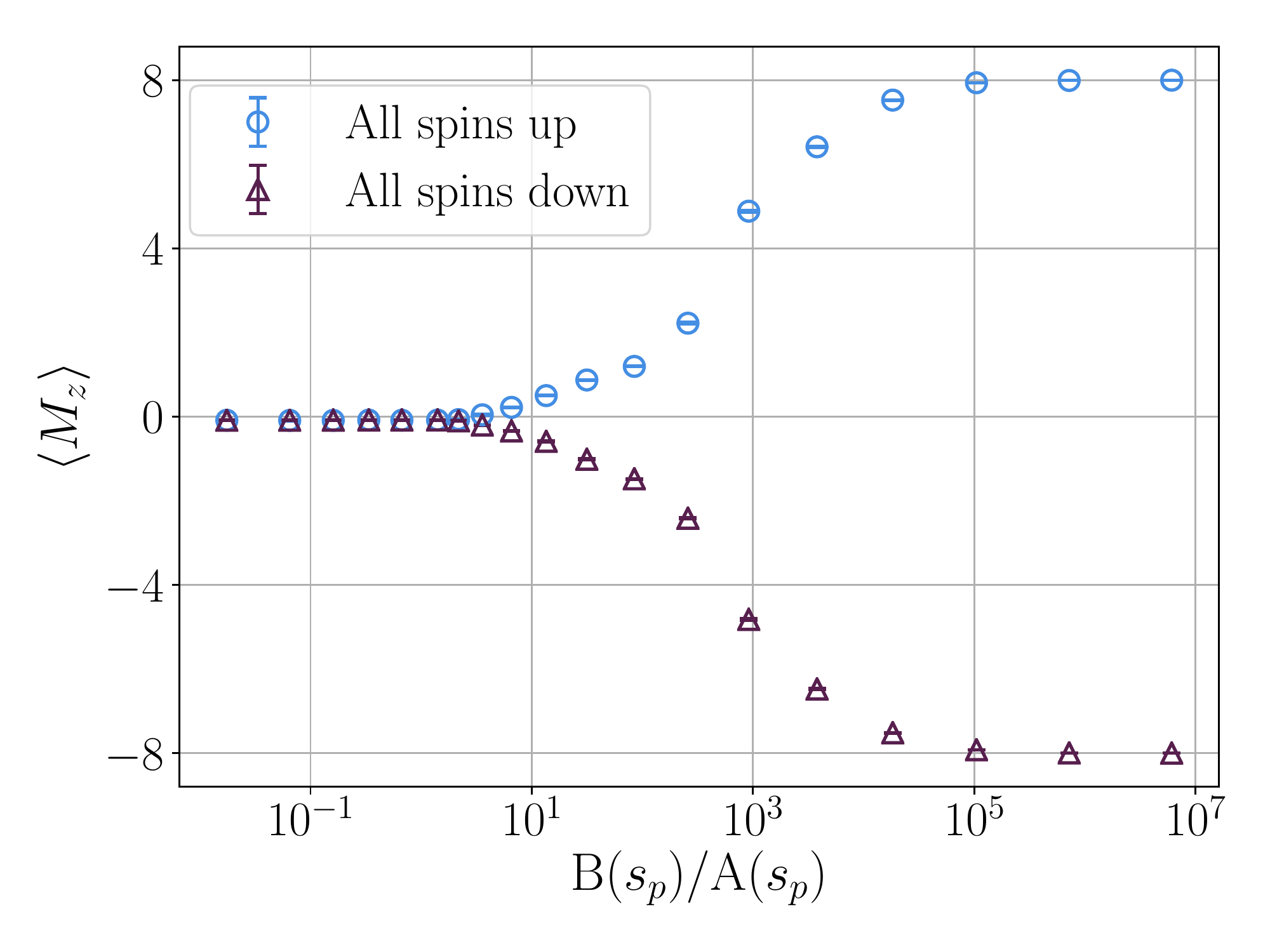}
\caption{Single-cell instances: Total mean magnetization $\langle M_z \rangle$ for local spins at $s=1$  
for disorder strength $\mathrm{B}(s_p)/\mathrm{A}(s_p)$, with $s_p$ denoting the pause dimensionless time. The initial states 
are taken as the all-up and all-down ferromagnetic states.}
\label{f5}
\end{figure}

\begin{figure}[!h]
\includegraphics[scale=0.38]{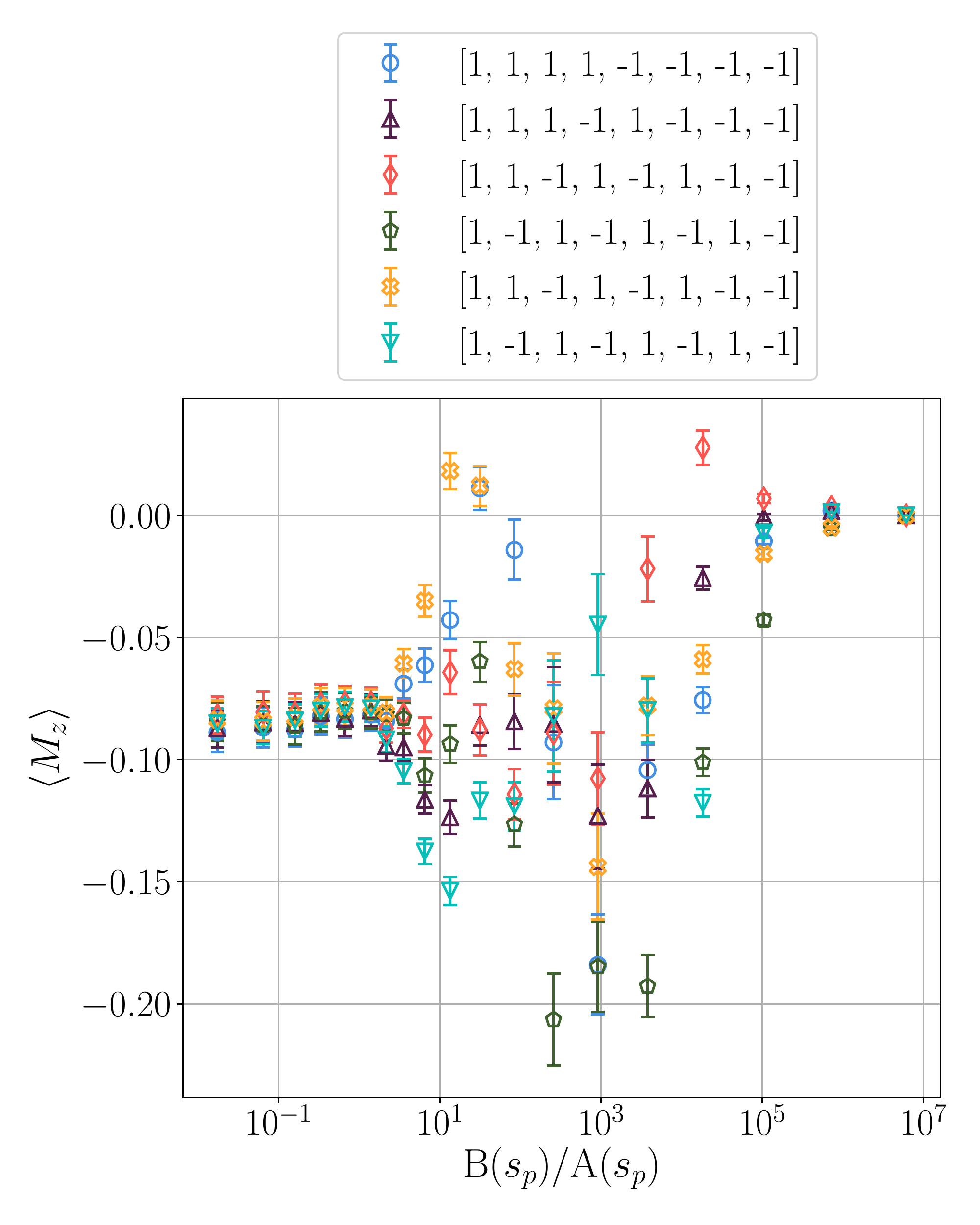}
\caption{Single-cell instances: Total mean magnetization $\langle M_z \rangle$ for local spins at $s=1$  
for disorder strength $\mathrm{B}(s_p)/\mathrm{A}(s_p)$, with $s_p$ denoting the pause dimensionless time. The initial states, all with 
zero magnetization, are shown in the legend.}
\label{f6}
\end{figure}

%%%%%%%%%%%%%%%%%%
\subsection{Single-cell instances}
%%%%%%%%%%%%%%%%%%
For single-cell runs, we use 48 different cells of the D-Wave processor, avoiding those on the edges of the processor and also
leaving at least one unused cell between them. On each of these cells, the same 200 random instances are
run, as described in Sec.~\ref{sec:tfim} and we set $J=1$.
To obtain each data point, we first perform a bootstrap over the final magnetization for the 
200 random instances, obtaining a mean local magnetization for each cell. Then we bootstrap over these 
individual cell mean magnetizations, and the mean obtained from that bootstrap constitutes a data point, with 
the error bars showing the $95\%$ confidence interval. Each data point corresponds to the magnetization at a 
specific disorder, defined as $\mathrm{B}(s_p)/\mathrm{A}(s_p)$, where $s_p$ is the pause location.
Each run consists of $1000$ anneals, each started at the initial state indicated in the plots. 
{The initial state here refers to the classical Ising eigenstates at the beginning of the reverse annealing process.} 

The results for initial states consisting of all spins up and all spins down are provided in Fig.~\ref{f5}, 
where we plot the total mean magnetization measured at $s=1$ as a function of the disorder strength. 
We observe that magnetization is found at zero value up to a limit of disorder strength, after which 
a memory effect can be observed. Remarkably, the onset of non-vanishing magnetization 
occurs in the range $\mathrm{B}(s_p)/\mathrm{A}(s_p) \in [2.5, 3.8]$, which is broadly in agreement with the magnitude order 
expected for disorder in the localization transition, according to the entanglement theoretical analysis for $N=8$ spins. 
The ratio $\mathrm{B}(s_p)/\mathrm{A}(s_p) \in [2.5, 3.8]$ is less than the theoretical critical point $(J / \Delta)_c \approx 4.8$ predicted for 
the eigenstate in the middle of the spectrum (the hardest to localize), as shown by the mobility edge for the energy phase diagram in Fig.~\ref{f3}.

We also test initial states with zero magnetization (Fig.~\ref{f6}) to confirm that the final magnetization remains close to
zero regardless of disorder, and find this to be the case, although there are small deviations from zero starting at a disorder that is again compatible with the 
localization transition. This greatly improves upon the higher-noise processor that we also tested (see Appendix~\ref{high_noise}), which presents much 
stronger fluctuations dependent on the specific initial state.

We also observe a slight bias towards $\langle M_z \rangle < 0$ for small disorder. One possible cause for this bias is spin-bath polarization~\cite{dw_manual}, 
which can sometimes occur during the QA process when consecutive anneals are performed without enough time between them, and the current 
continuously flowing through the qubits polarizes the environment leading to correlation between samples. However, we do not observe any significant differences 
after varying the time between anneals, making it unlikely for spin-bath polarization to be the culprit. A global local field bias that favors negative magnetization has 
also been recently reported in a distinct D-Wave device~\cite{Nelson:21}.

%%%%%%%%%%%%%%%%%%
\subsection{Double-cell instances}
\label{2cell}
%%%%%%%%%%%%%%%%%%
For double-cell runs, we use 34 different pairs of cells on the D-Wave processor, with the 2 cells of the pair horizontally connected, 
as represented in Fig.~\ref{f1}. The cells on the edges of the processor are avoided and at least one unused cell is left 
between each pair. 

\begin{figure}[!hb]
\centering
\includegraphics[scale=0.4]{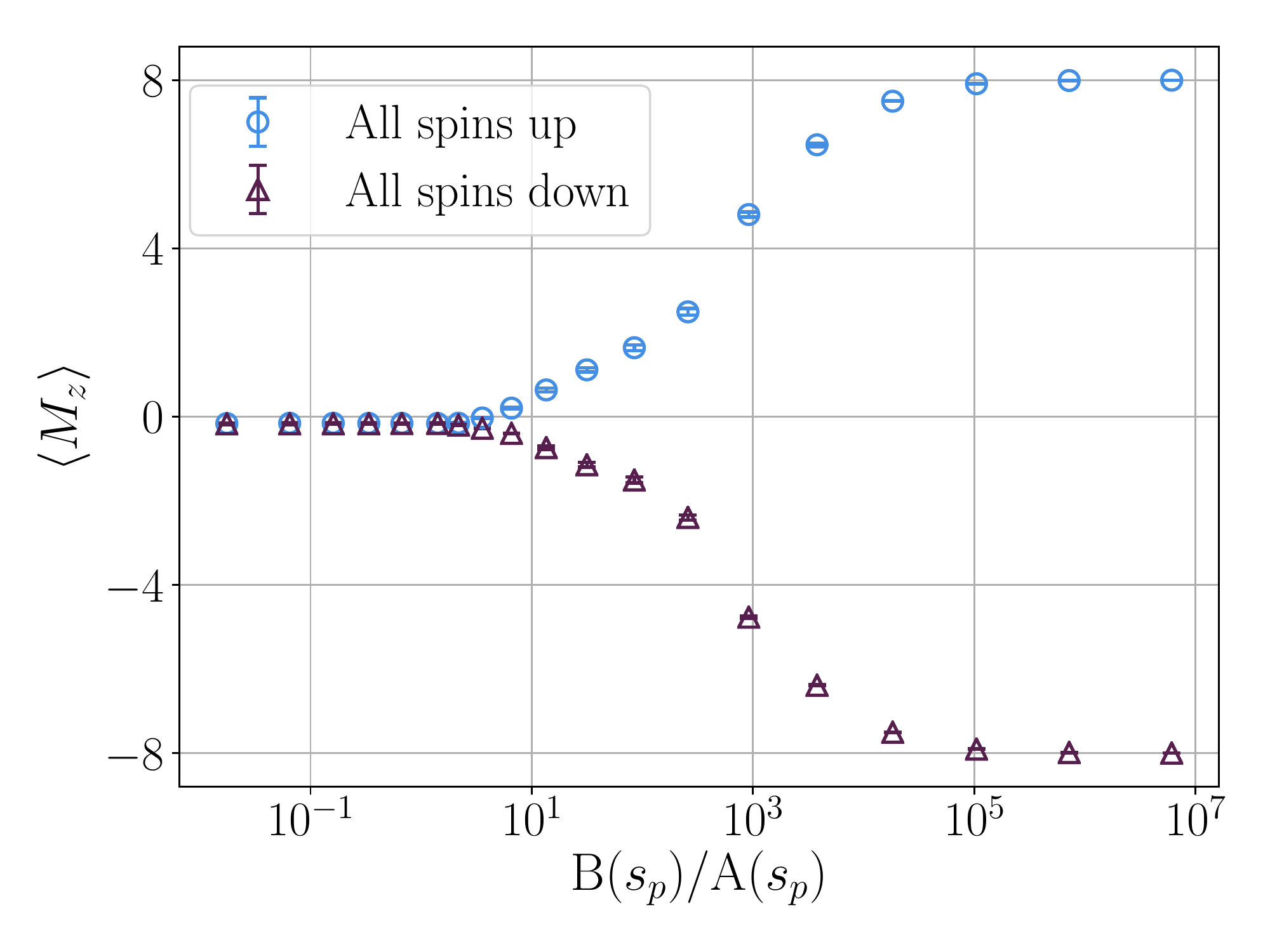}
\caption{Double-cell instances: Total mean magnetization $\langle M_z \rangle$ for local spins at $s=1$  
for disorder strength $B(s_p)/A(s_p)$, with $s_p$ denoting the pause dimensionless time. The initial states 
are taken as the all-up and all-down ferromagnetic states.}
\label{f8}
\end{figure}

We run 200 random instances, with the data points and error bars calculated as in the previous case.
The results for initial states consisting of all spins up and all spins down are provided in Fig.~\ref{f8}. 
Similar to the single-cell instances, the final magnetization is zero up to a limit of disorder strength, after which 
a memory effect can be observed. The onset of a non-vanishing magnetization 
also occurs at $\mathrm{B}(s_p)/\mathrm{A}(s_p) \in [2.5, 3.8]$, which indicates robustness of the critical point against larger instances. 

The initial states with zero magnetization show very similar behavior to those in the single-cell case, with small fluctuations in $\langle M_z \rangle$ for the same region, 
which is compatible with the localization transition. The preference for $\langle M_z \rangle < 0$, rather than $\langle M_z \rangle=0$ at low disorder also remains. 
Results for a few of these states, comparing the two different processors, can be found in Appendix~\ref{high_noise}.

%%%%%%%%%%%%%%%%%%
\subsection{Disorder and memory effects in the D-Wave experiment}
\label{verification}
%%%%%%%%%%%%%%%%%%

We have so far confirmed that the onset of memory effects in the D-Wave experiments occurs at a disorder compatible with that predicted by theory. 
This can be taken as an indication of the disorder-induced localized phase in the D-Wave chip. To strengthen the validity of our observations, we have run 
additional experiments to rule out that the observed memory effects could emerge for causes other than disorder.

To this end, we choose a uniform antiferromagnetic system on a single Chimera cell, with the same connectivity as our original disordered system 
(i.e. a fully connected cell), but all $J_{ij}=1$ instead of randomly chosen from $[-1, 1]$. Rather than the all spins up (or all spins down) initial state, 
we set it to be $[1, 1, 1, 1, 1, 1, 1, -1]$, the reason being that the all spins up or down states will on average be in the middle of the energy spectrum 
for the disordered cell, while they would instead be at the top for the uniform case. The state $[1, 1, 1, 1, 1, 1, 1, -1]$ avoids the top of the spectrum 
for the uniform system, providing a fairer comparison. Note that the magnetizations will then be different, which we expect to see at high values of $\mathrm{B}(s_p)/\mathrm{A}(s_p)$. 

\begin{figure}[!hbt]
\centering
\includegraphics[scale=0.4]{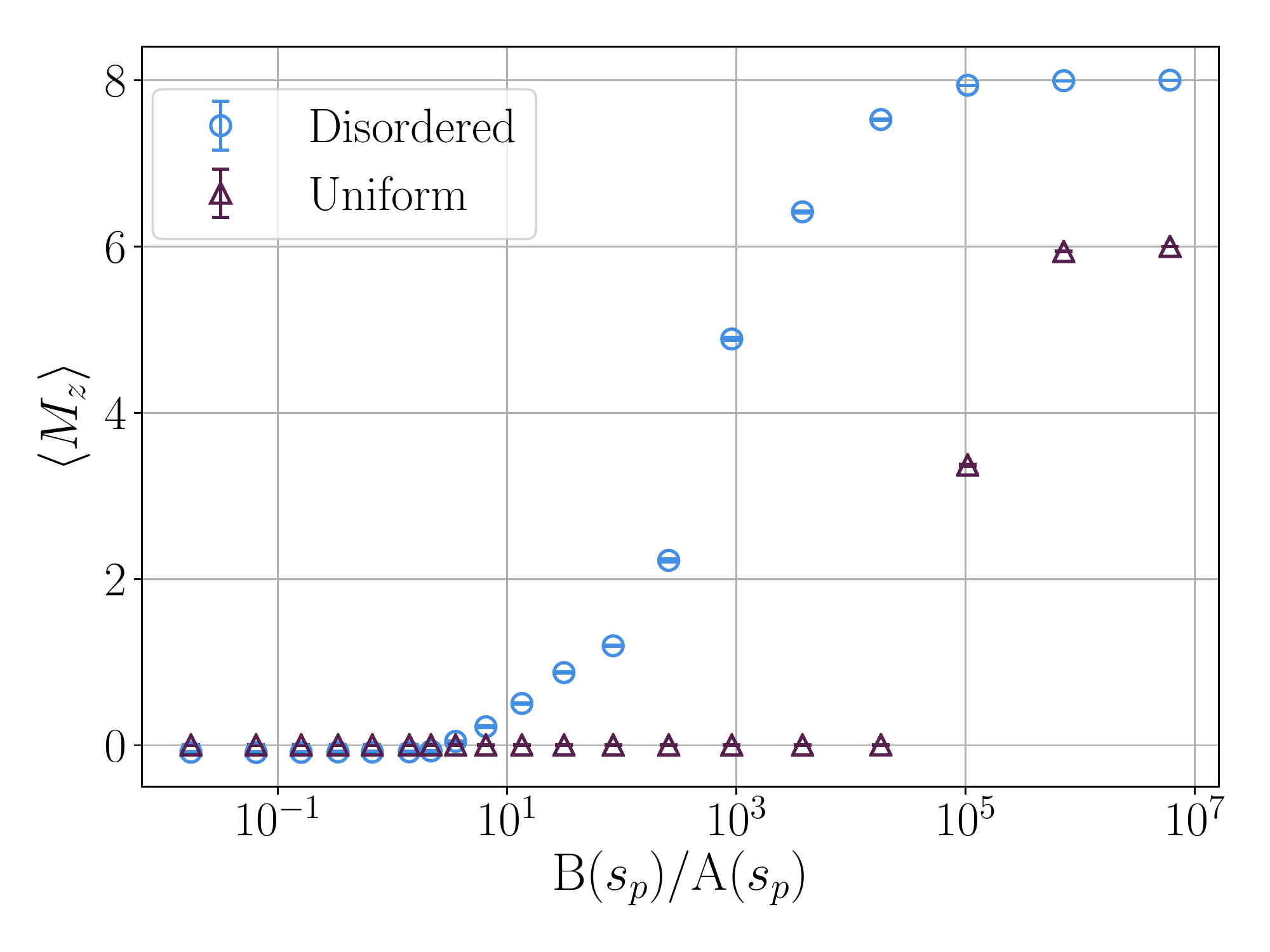}
\caption{Comparison to non-disordered system: Total mean magnetization $\langle M_z \rangle$ for local spins at $s=1$  
as a function of $\mathrm{B}(s_p)/\mathrm{A}(s_p)$, with $s_p$ denoting the pause dimensionless time. The initial state
is taken as the all-up ferromagnetic states for the disordered cell, and $[1, 1, 1, 1, 1, 1, 1, -1]$ for the antiferromagnetic uniform cell.}
\label{f9}
\end{figure}

What we wish to compare is the location at which memory effects start to appear.
Figure~\ref{f9} shows this comparison: for the uniform, non-disordered cell, we do not observe a deviation from $\langle M_z \rangle = 0$ until $\mathrm{B}(s_p)/\mathrm{A}(s_p) \approx 10^5$, 
several orders of magnitude later than for the disordered one. The memory effect is in this case unrelated to the localized phase, and simply corresponds to the fact that, when the 
reverse annealing is performed to a very late $s_p$, the transverse field is too weak to make the system leave its initial state.

\begin{figure}[!hbt]
\centering
\includegraphics[scale=0.395]{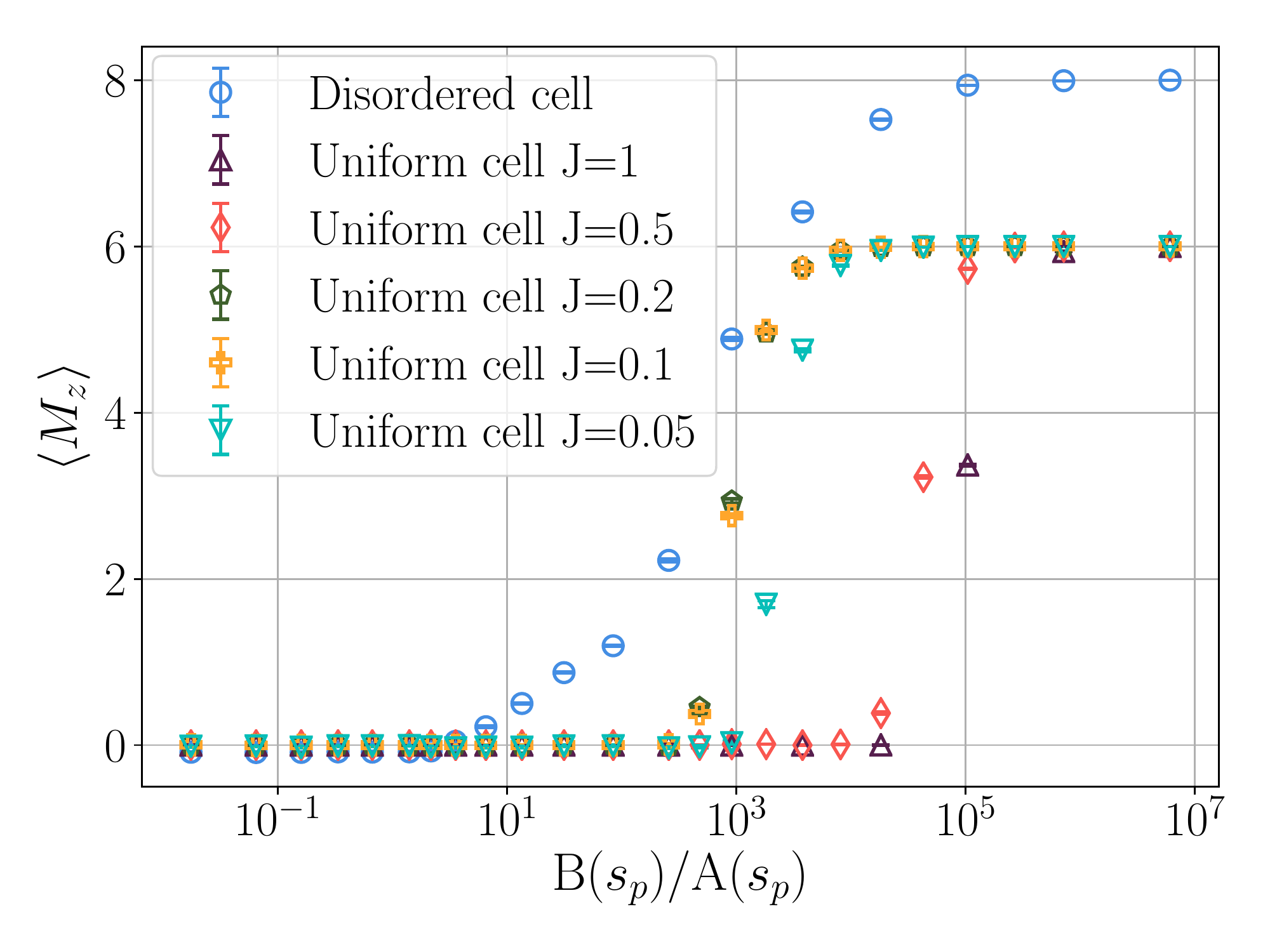}
\caption{Comparison to non-disordered system at different energy scales: Total mean magnetization $\langle M_z \rangle$ for local spins at $s=1$  
as a function of $\mathrm{B}(s_p)/\mathrm{A}(s_p)$, with $s_p$ denoting the pause dimensionless time. The initial state
is taken as the all-up ferromagnetic states for the disordered cell, and $[1, 1, 1, 1, 1, 1, 1, -1]$ for the uniform cell. The coupling strength for the uniform cell is indicated in the legend.}
\label{f10}
\end{figure}

We also consider the effect that the energy scale might have on the onset of memory effects unrelated with localization. For this, we repeat the experiment on the uniform cell, with 
different values of $J$ (the uniform coupling strength). As we decrease $J$, the appearance of the memory effects moves to smaller $\mathrm{B}(s_p)/\mathrm{A}(s_p)$, but only up to a point; once $J$ 
decreases below 0.1, the direction shifts and memory appears at larger $\mathrm{B}(s_p)/\mathrm{A}(s_p)$. This is portrayed in Fig.~\ref{f10}. $J=0.1$ and $J=0.2$ lead to the earliest onset of memory 
effects for the uniform cell, around $\mathrm{B}(s_p)/\mathrm{A}(s_p) \approx 258$, still very far from the location when disorder is present.

%%%%%%%%%%%%%%%%%%
\subsection{Classical simulation of the annealing process}
\label{SVMC}
%%%%%%%%%%%%%%%%%%

D-Wave results might suggest we are experimentally observing the theoretically predicted localization transition. However, at this stage, we can assert only that compatible results with 
transitions has been obtained. This is because we can show that the spin-vector Monte Carlo (SVMC) approach -- a classical Monte Carlo simulation method whereby qubits are replaced by 2-dimensional rotors-- is able to reproduce the 
experimental results. SVMC does not simulate entanglement, yet it has had a lot of success reproducing the qualitative features of the output statistics of the D-Wave quantum annealers \cite{SSSV:14,Hen:15,Albash:15,Albash:15b,Mishra:18,Li:20,Albash:21}. It thus remains an important algorithm to compare against in order to ascertain whether quantum effects are important in determining the output statistics. Our implementation of SVMC is modified \cite{Albash:21} to better capture the slow thermal dynamics at large $s$ \cite{Amin:15}, and we give details of the algorithm in Appendix~\ref{app:SVMC}.

We show in Fig.~\ref{fig:SVMC} results using our implementation of SVMC.  The simulation reproduces the key features of Fig.~\ref{f10}, specifically that the magnetization of the disordered cell deviates from $0$ at significantly earlier values of $\mathrm{B}(s_p)/\mathrm{A}(s_p)$ compared to the uniform cell.  The SVMC simulations saturate at smaller values of $\mathrm{B}(s_p)/\mathrm{A}(s_p)$ than the experimental results, but we expect that further tuning of the simulation dynamics would help get a better quantitative agreement.  Nonetheless, it is clear that this purely classical model qualitatively reproduces the memory effects obtained through the quantum annealer as signatures of the MBL transition.

\begin{figure}[!hbt]
\centering
\includegraphics[scale=0.39]{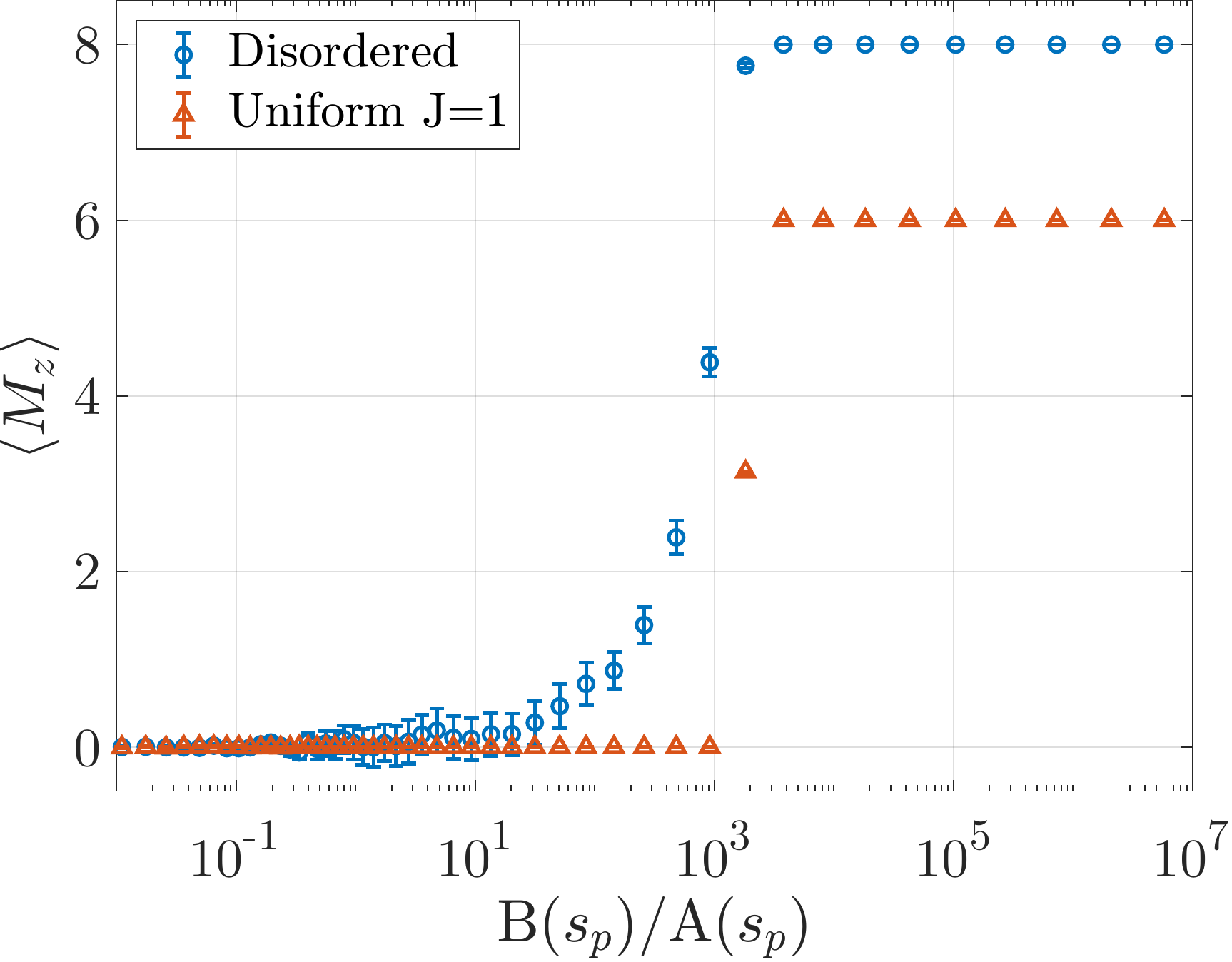}
\caption{Comparison of the disordered and non-disordered systems at different energy scales using SVMC: Total mean magnetization $\langle M_z \rangle$ for local spins at $s=1$ as a function of $\mathrm{B}(s_p)/\mathrm{A}(s_p)$, with $s_p$ denoting the pause dimensionless time. The initial state is taken as the all-up ferromagnetic states for the disordered cell, and $[1, 1, 1, 1, 1, 1, 1, -1]$ for the uniform cell. The coupling strength for the uniform cell is indicated in the legend.  Simulation parameters are given in Appendix~\ref{app:SVMC}.}
\label{fig:SVMC}
\end{figure}

%%%%%%%%%%%%%
\section{Summary and conclusions}
%%%%%%%%%%%%%

We reported on an investigation of a localization phase transition in the spin-1/2 transverse-field Ising model defined 
on a Chimera connectivity graph .
We detected the critical point using the variance of the block entanglement. We also found that the usage of mean block 
entanglement and energy level statistics explicitly posed the localization phase transition 
as a dynamical transition rooted in the individual behavior of the energy eigenstates. 

We then devised and ran experiments on two DWk2Q processors using a combination of the reverse annealing technique with the pause-quench protocol 
to locate the critical point associated with this localization phase transition via local magnetization measurements. The results 
obtained were shown to be consistent with our theoretical predictions. 
We also demonstrated that SVMC, a classical model of the system, reproduces the same experimental signature. Thus, we emphasize that the mean local magnetization, as measured in our work,  does not provide a purely quantum signature of the phase transition in the DW2kQ, even though the critical point obtained is 
compatible with the theoretical prediction. 
{The investigation of local observables and larger system sizes providing results beyond the SVMC classical model are topics left for future research.}

Many-body localized systems exhibit a fascinating interplay between interaction and disorder. 
Their simulation in fully controllable devices provide a general mechanism to approach 
the separation between extended states and states localized by disorder. Although the Chimera connectivity has been 
explicitly considered, our characterization can be adapted to a variety of lattice topologies within the QA framework.  
We believe that the setup proposed here is potentially fruitful for further investigations 
and experimental realizations of general critical properties of disordered Ising systems. 

\begin{acknowledgments}
We thank Evgeny Mozgunov for helpful discussions. 
The research is based upon work (partially) supported by the Office of
the Director of National Intelligence (ODNI), Intelligence Advanced
Research Projects Activity (IARPA) and the Defense Advanced Research Projects Agency (DARPA), via the U.S. Army Research Office
contract W911NF-17-C-0050. The views and conclusions contained herein are
those of the authors and should not be interpreted as necessarily
representing the official policies or endorsements, either expressed or
implied, of the ODNI, IARPA, DARPA, or the U.S. Government. The U.S. Government
is authorized to reproduce and distribute reprints for Governmental
purposes notwithstanding any copyright annotation thereon.
Z.G.I. was  supported  by  NASA Academic  Mission  Services  (NAMS), contract  number NNA16BD14C, as well as by the AFRL Information Directorate under  grant  F4HBKC4162G001  and  the  Office  of  the Director of National Intelligence (ODNI) and the Intelligence  Advanced  Research  Projects  Activity  (IARPA), via IAA 145483.
M.S.S. acknowledges financial support from the Conselho Nacional de Desenvolvimento  Científico e Tecnológico  (CNPq) (No. 307854/2020-5).  
This research is also supported in part by Coordena\c{c}\~ao de Aperfei\c{c}oamento de Pessoal de N\'{\i}vel Superior  - Brasil (CAPES) 
(Finance Code 001) and by the Brazilian National Institute for Science and Technology of Quantum Information [CNPq INCT-IQ (465469/2014-0)]. 
This material is also based upon work supported by the National Science Foundation the Quantum Leap Big Idea under Grant No. OMA-1936388.
\end{acknowledgments}

\appendix

%%%%%%%%%%%%%%%%%%
\section{Entanglement and bipartitions of the unit Chimera cell}
\label{LR_partition}
%%%%%%%%%%%%%%%%%%

\begin{figure}[!hbt]
\includegraphics[scale=0.245]{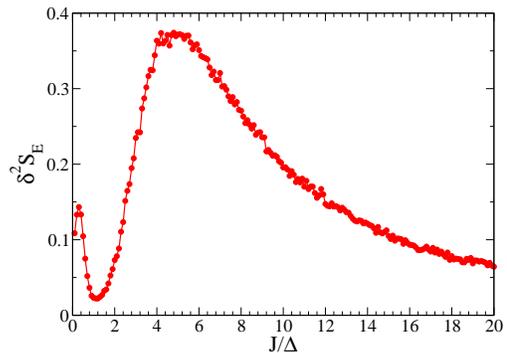}
\caption{Variance of the mean block entanglement for a left-right partition as a function of the disorder strength. Its maximum value, which is the 
precursor of the critical point, occurs at $(J / \Delta)_c \approx 4.8$. }
\label{f-ap}
\end{figure}

We can show that the characterization of the critical point through the mean block entanglement can be achieved through different bipartitions of the system. 
We focus on $N=8$ spins, since this is size of the unit Chimera cell implemented on the DW2kQ. Instead of an up-down partition, such as previously implemented, 
we consider a left-right cut in the Chimera cell, splitting out the spins in two subsets given by $A = \{0,1,2,3\}$ and $B = \{4,5,6,7\}$, 
as labeled in Fig.~\ref{f1}. Our analysis is again carried out for the eigenstate in the middle of the energy spectrum. 
We perform averages over $5\times10^3$ disorder configurations. We then 
evaluate the variance of the mean block entanglement for disorder ensembles as a function 
of the disorder strength. The results are shown in Fig.~\ref{f-ap}. Notice that the maximum of the variance 
occurs at $(J / \Delta)_c \approx 4.8$, which is in agreement with the result previously obtained for the up-down partition. 
\begin{figure}[!bt]
\centering
\includegraphics[scale=0.305]{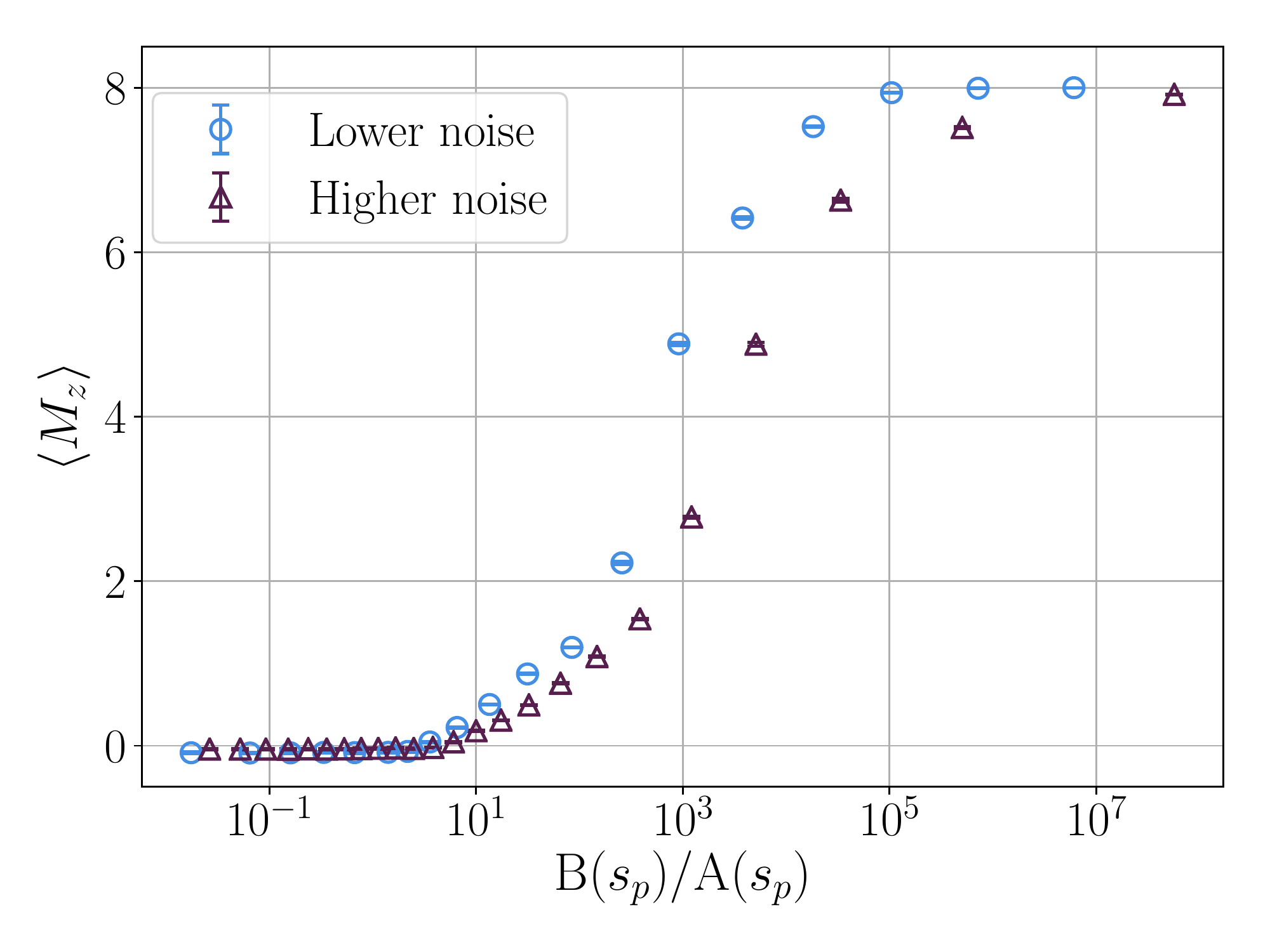}
\includegraphics[scale=0.305]{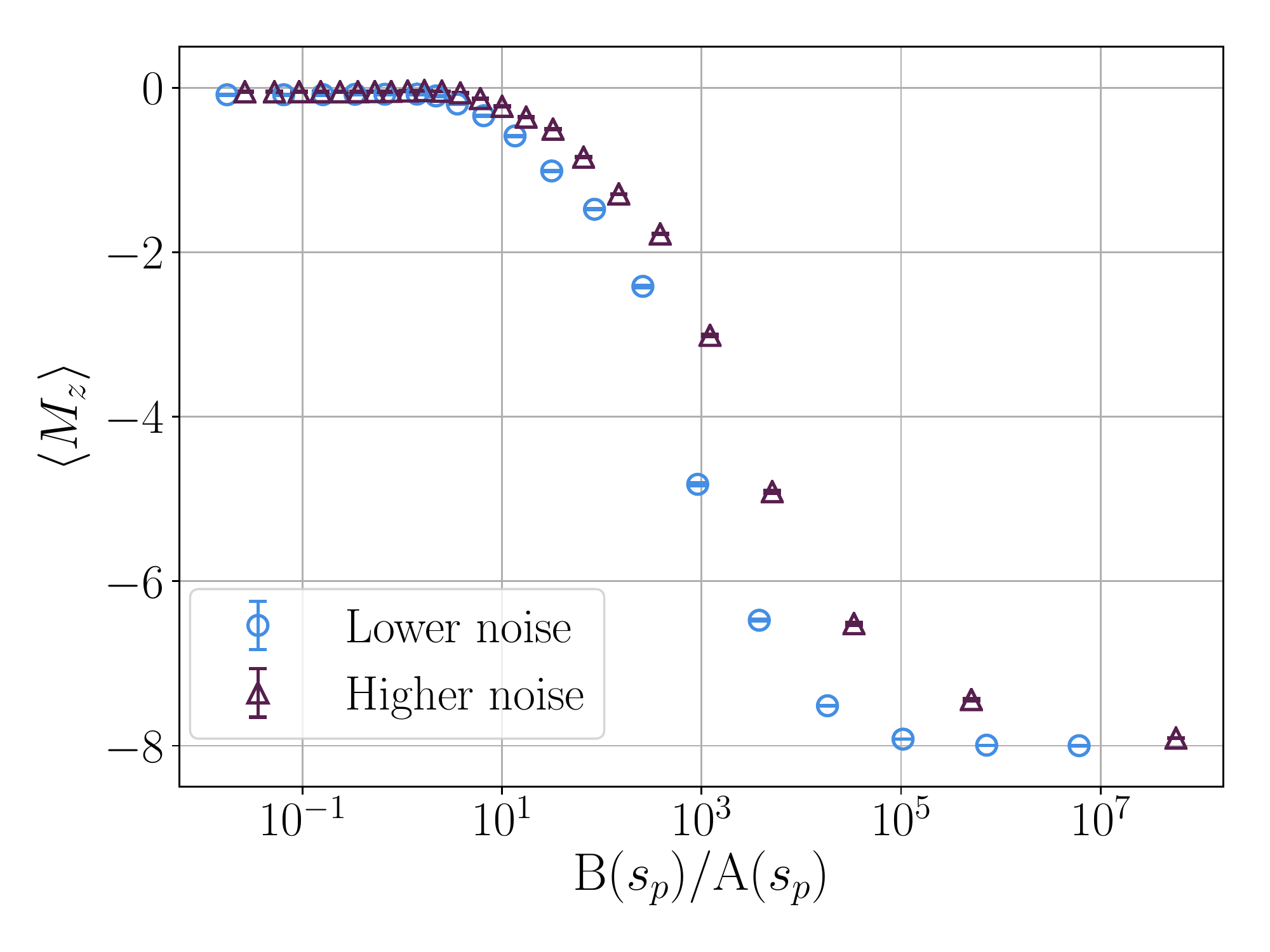}
\caption{Comparison between baseline and lower noise processor: Total mean magnetization for single-cell instances, for the all-up and all-down ferromagnetic states.}
\label{f11}
\end{figure}
\begin{figure}[!h]
\centering
\includegraphics[scale=0.305]{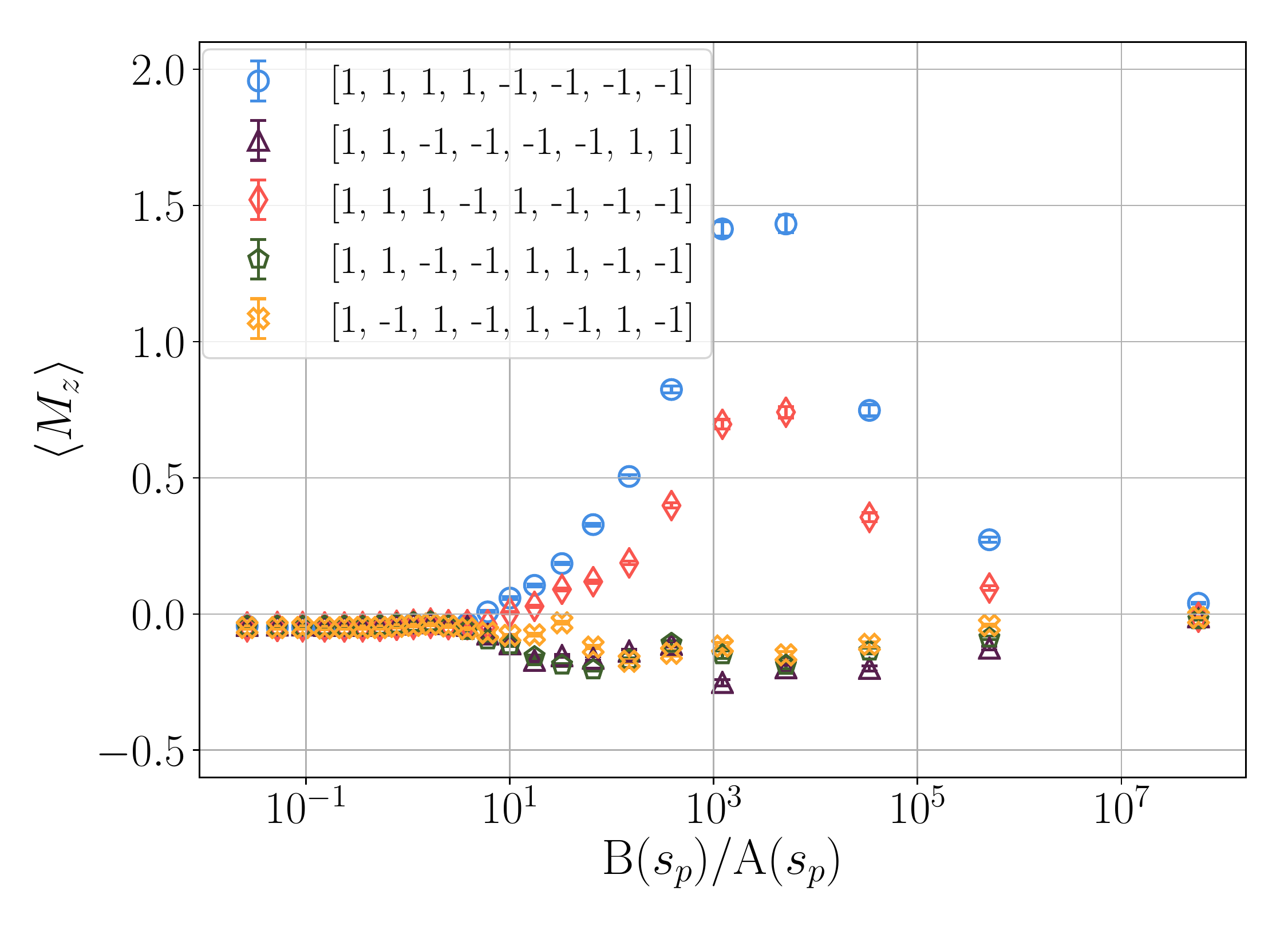}
\includegraphics[scale=0.305]{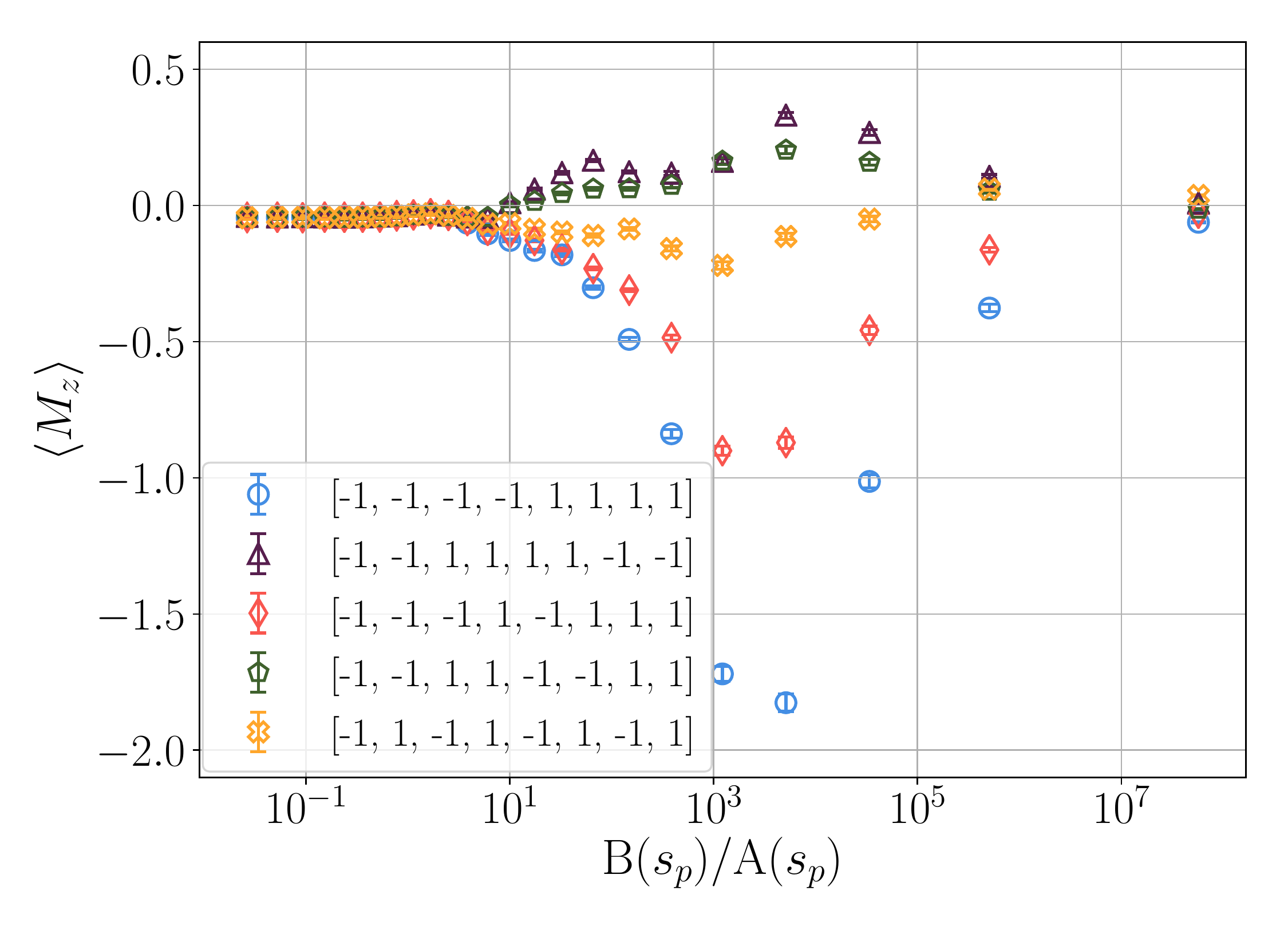}
\caption{Single-cell instances: Total mean magnetization $\langle M_z \rangle$ for zero magnetization initial states, shown in the legend.}
\label{f12}
\end{figure}

\begin{figure}[!hbt]
\centering
\includegraphics[scale=0.34]{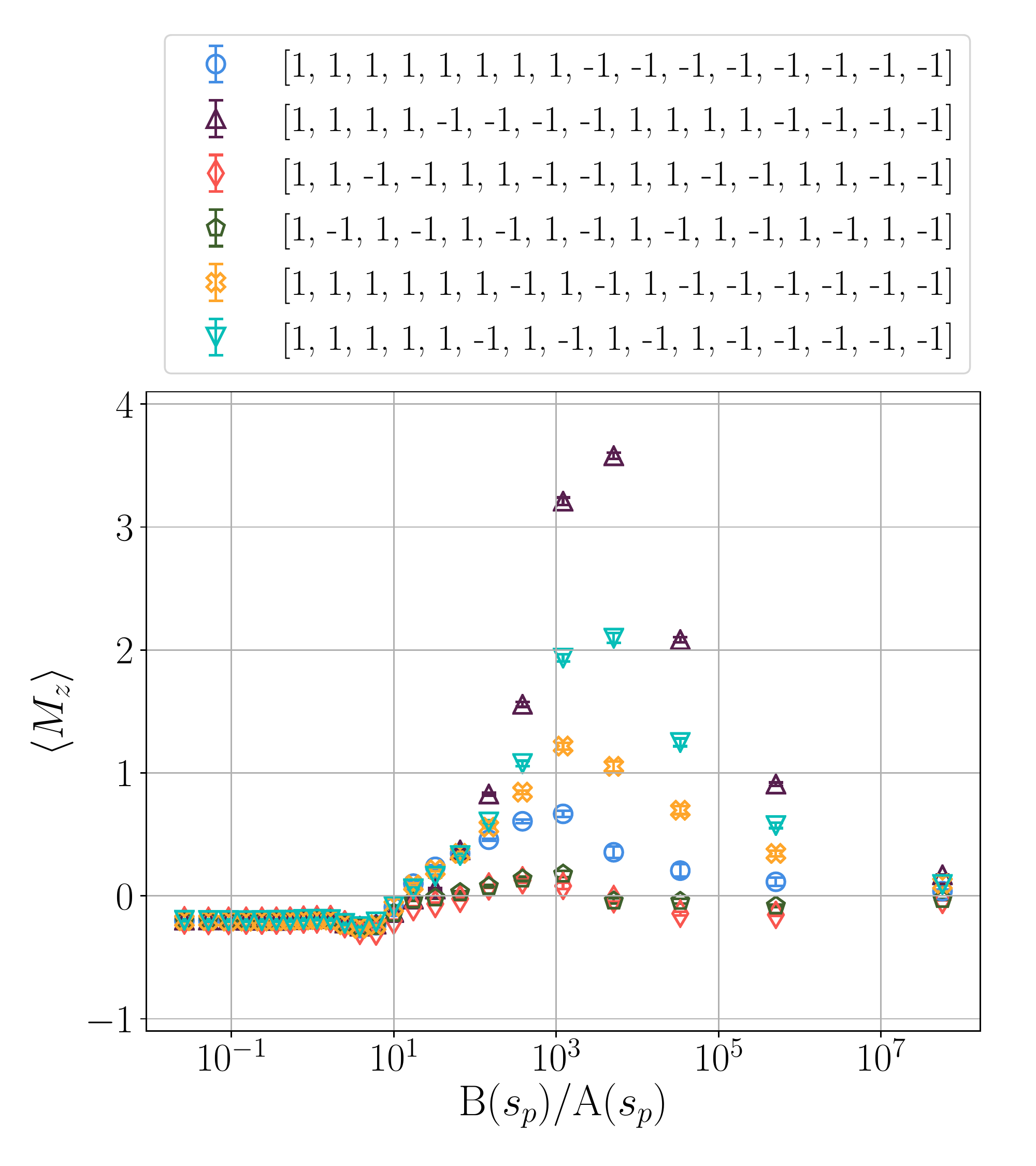}
\caption{Double-cell instances: Total mean magnetization $\langle M_z \rangle$ for local spins at $s=1$  
for disorder strength $\mathrm{B}(s_p)/\mathrm{A}(s_p)$, with $s_p$ denoting the pause dimensionless time. The initial states 
are taken as some zero magnetization configurations.}
\label{f13}
\end{figure}

%%%%%%%%%%%%%%%%%%
\section{Comparison with higher-noise device}
\label{high_noise}
%%%%%%%%%%%%%%%%%%

We present additional results obtained from a previous version of the DW2kQ, housed at NASA Ames Research Center. Due to improvements 
in the fabrication process, it is noisier than the one used in the main text. We still consider it important to report these results, which illustrate 
some of the perils of noisy quantum devices.

As a starting point, Fig.~\ref{f11} shows that the results for the all spins up and all spins down initial states are still compatible with the phase transition, and very similar to those of the lower noise 
processor as we would expect, although the lower noise processor recovers the full memory of the initial state earlier. The same behavior is observed for the double-cell instances.

For the initial states with zero magnetization, however, we see significant differences between the two processors. We find, as shown in Fig.~\ref{f12}, that for the higher noise 
device, the final magnetization varies greatly depending on the specific initial state. 
The behavior of pairs of initial states with spins flipped with respect to each other is approximately symmetrical, although a slight bias to $\langle M_z \rangle < 0$ is observed throughout.
Initial states in which biases are concentrated in the middle (such as $[1, 1, 1, 1, -1, -1, -1, -1]$) appear to show larger fluctuations in $\langle M_z \rangle$ than those with more staggered 
biases (e.g., $[1, -1, 1, -1, 1, -1, 1, -1]$).
We repeat the experiment on two-cell instances, to account for the possibility that the fluctuations could be due to finite size effects (see Fig.~\ref{f13}), but the permanence of these fluctuations and their scaling as we increase 
the size of the system makes this unlikely, and points instead to
noise mechanisms present in the hardware regardless of instance size.

Just like the single-cell instances, the deviation from zero magnetization shows a dependency on how staggered
the biases are in the initial state. 
In particular, initial states with locally staggered 
magnetization concentrated in the middle of the system are the ones that deviate the most from zero.
For instance, the configuration $[1, 1, 1, 1, -1, -1, -1, -1, 1, 1, 1, 1, -1, -1, -1, -1]$ exhibits the strongest fluctuations 
after the critical point, similar to $[1, 1, 1, 1, -1, -1, -1, -1]$ (same pattern on a single cell) did for the single-cell case.
Finally, we compare these results to those obtained from the lower noise device that were presented in the main text. Fig.~\ref{f14} and Fig.~\ref{f15} show how the large fluctuations even out with the noise reduction, although the preference for $\langle M_z \rangle < 0$ is slightly more marked in the lower noise processor.

%%%%%%%%%%%%%%%%%%
\section{Spin-Vector Monte Carlo}
\label{app:SVMC}
%%%%%%%%%%%%%%%%%%

In Spin-Vector Monte Carlo (SVMC) \cite{SSSV:14}, the dynamics of a noisy quantum annealer are approximated using a system of rotors with a time-dependent energy potential.  In this model, each qubit of the $n$-qubit system is replaced by a 2-dimensional rotor characterized by an angle $\theta \in [0, 2\pi)$, and under the mapping $\sigma_i^z \to \cos \theta_i, \sigma_i^x \to \sin \theta_i$ the time-dependent quantum Hamiltonian (Eq.~\eqref{eqt:DWHamil}) gives the energy potential for the system of rotors:
\begin{eqnarray}
V(s, {\theta}) &=& - A(s) \sum_{i=1}^n \sin \theta_i \nonumber \\
&& \hspace{-0.5cm} + B(s) \left( \sum_{i=1}^n \cos \theta_i + \sum_{\langle i,j \rangle} J_{ij} \cos \theta_i \cos \theta_j \right) \ .
\end{eqnarray}
This potential can be understood as the energy potential that arises from the spin-coherent path-integral formalism \cite{Owerre:15,Albash:15,Albash:15b} , and the restriction to a 2-dimensional rotor as opposed to a normalized 3-dimensional vector can be understood as arising from the strong coupling to a thermal environment \cite{Cro2016}.

\begin{figure}[!hbt]
\centering
\includegraphics[scale=0.34]{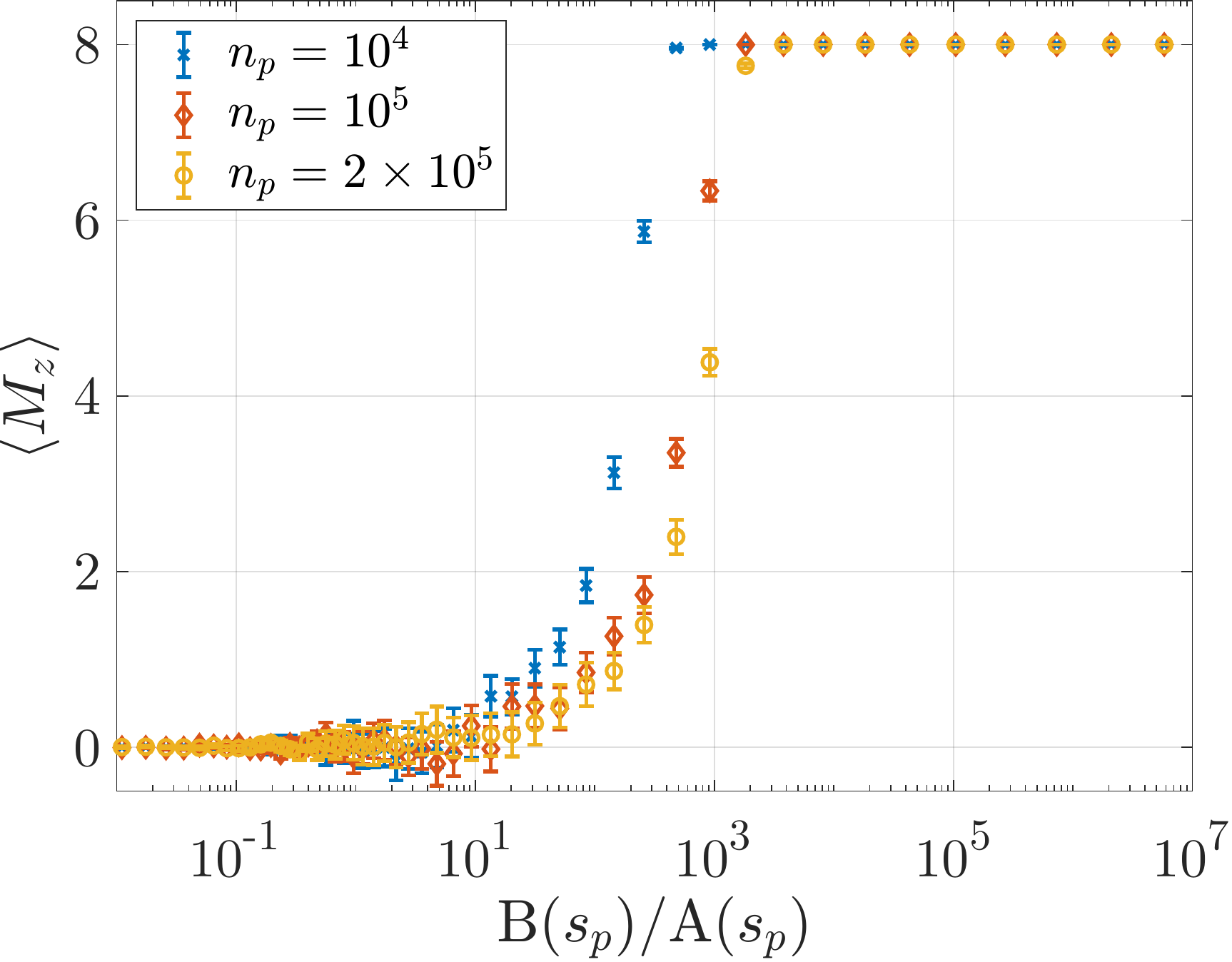}
\includegraphics[scale=0.34]{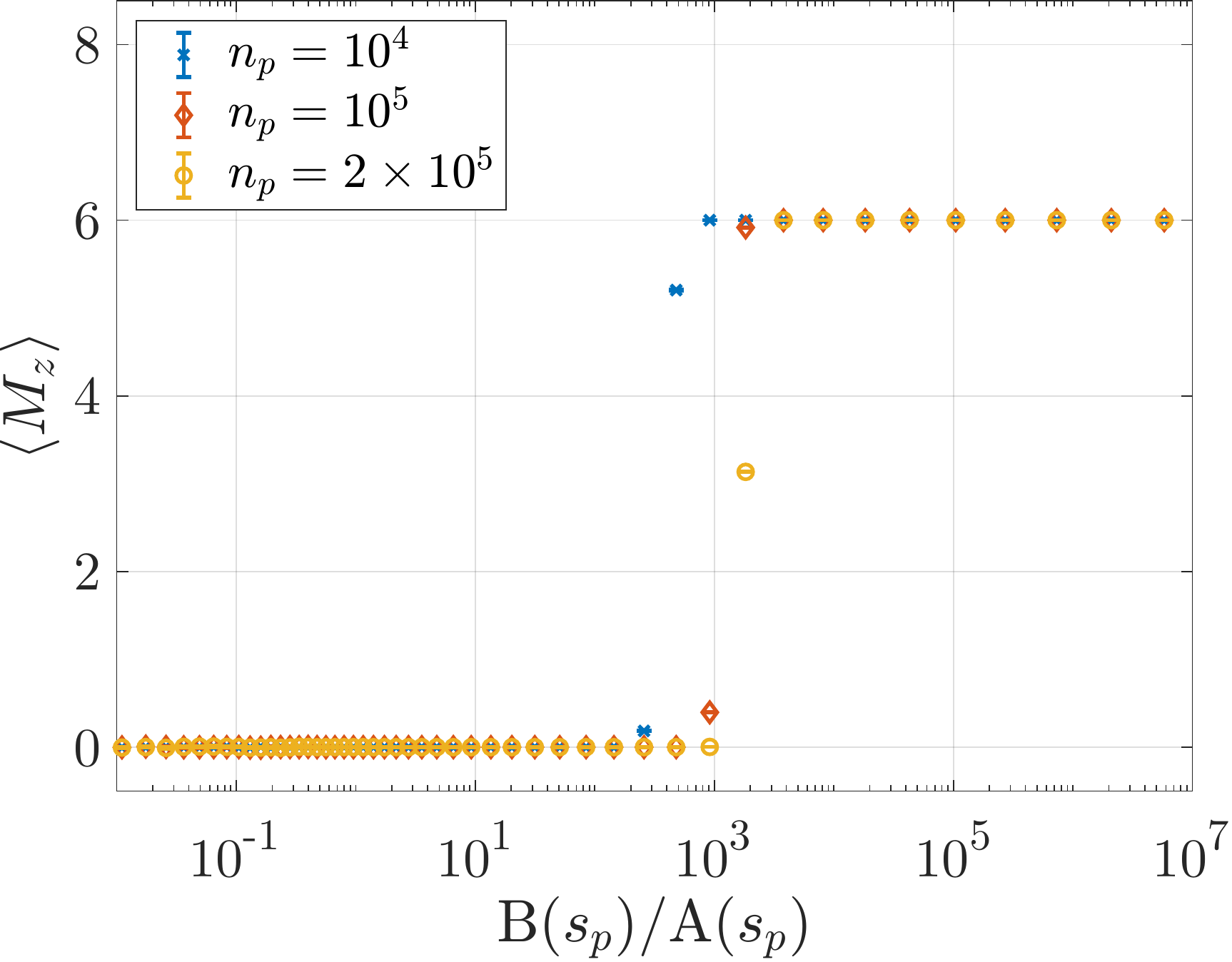}
\caption{Comparison between different number of SVMC pause sweeps $n_p$. Total mean magnetization $\langle M_z \rangle$ for disordered cells with the initial state $[1, 1, 1, 1, 1, 1, 1, 1]$ (top) and uniform cells with the initial state $[1, 1, 1, 1, 1, 1, 1, -1]$ (bottom).}
\label{fig:SVMC2}
\end{figure}

The system of rotors evolves using a Metropolis-Hastings algorithm \cite{Met1953,Has1970}: for each rotor, a new configuration $\theta_i'$ is randomly chosen in the range $[0, 2\pi)$ and accepted according to the Metropolis probability $p = \min \left( 1, \exp\left( - \beta \Delta V \right)\right)$ where $\Delta V$ is the change in potential energy if the new configuration is accepted and $\beta$ is a fixed inverse-temperature. We take the temperature to be given by $k_B T/\hbar = 1.57146 \mathrm{GHz}$, corresponding to a temperature of 12$m\mathrm{K}$. In our simulations, we use the same annealing protocol for $\mathrm{A}(s)$ and $\mathrm{B}(s)$ as used by the physical quantum annealer: (a) we rapidly change from $s = 1 \to s_\ast$; (b) we pause at $s_\ast$; (c) we rapidly change from $s = s_\ast \to 1$. For parts (a) and (c), we use $\lfloor 2 \times 10^3 \times (1-s_\ast) \rfloor$ sweeps (a sweep corresponds to proposing a new configuration for each rotor once), and then for part (b) we use a $2 \times 10^5$ sweeps.  We do not find significant qualitative differences as we change the total number of sweeps for part (b), as we show in Fig.~\ref{fig:SVMC2}.

\begin{widetext}

\begin{figure}[!hbt]
\centering
\includegraphics[scale=0.34]{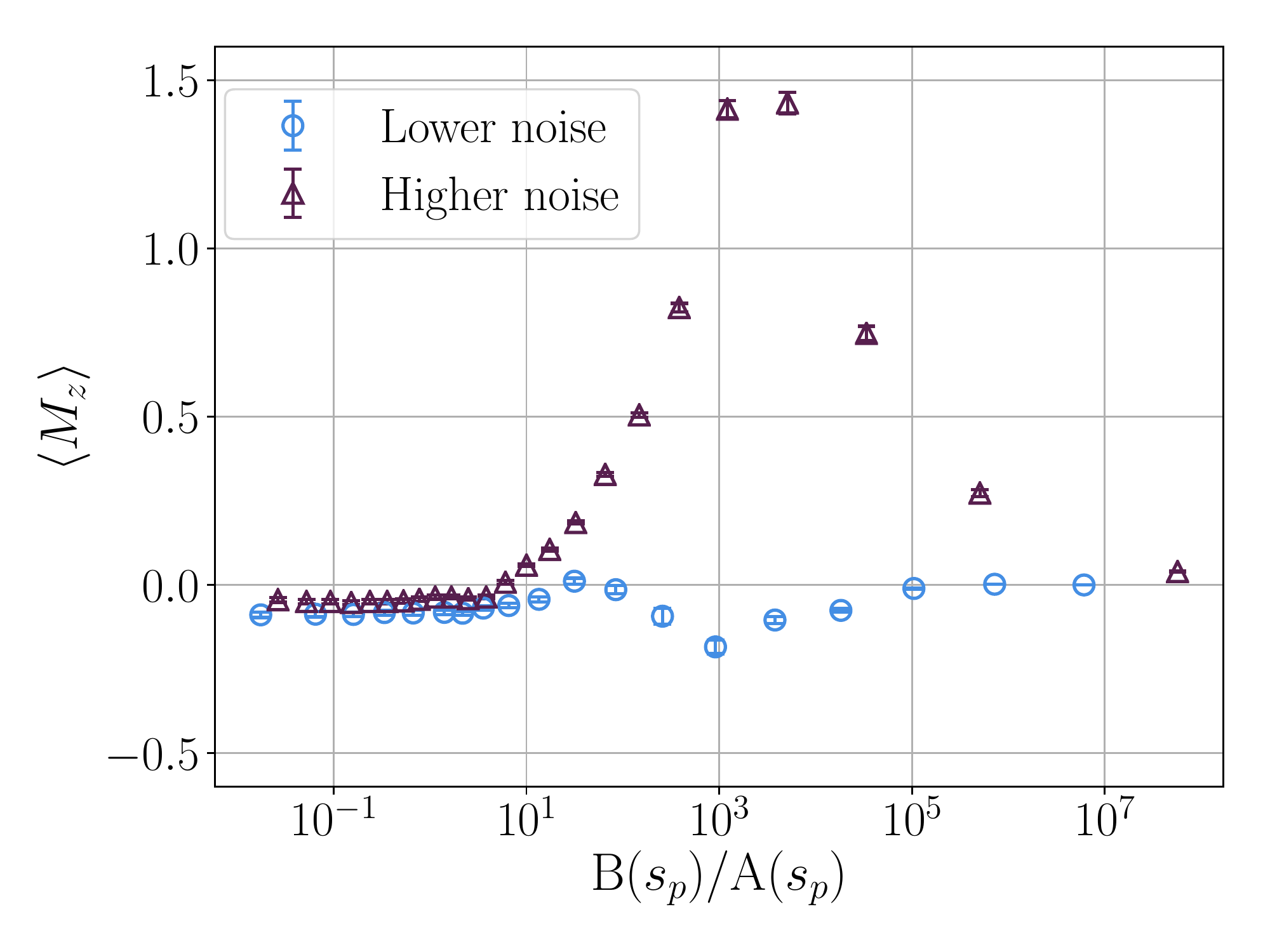}
\includegraphics[scale=0.34]{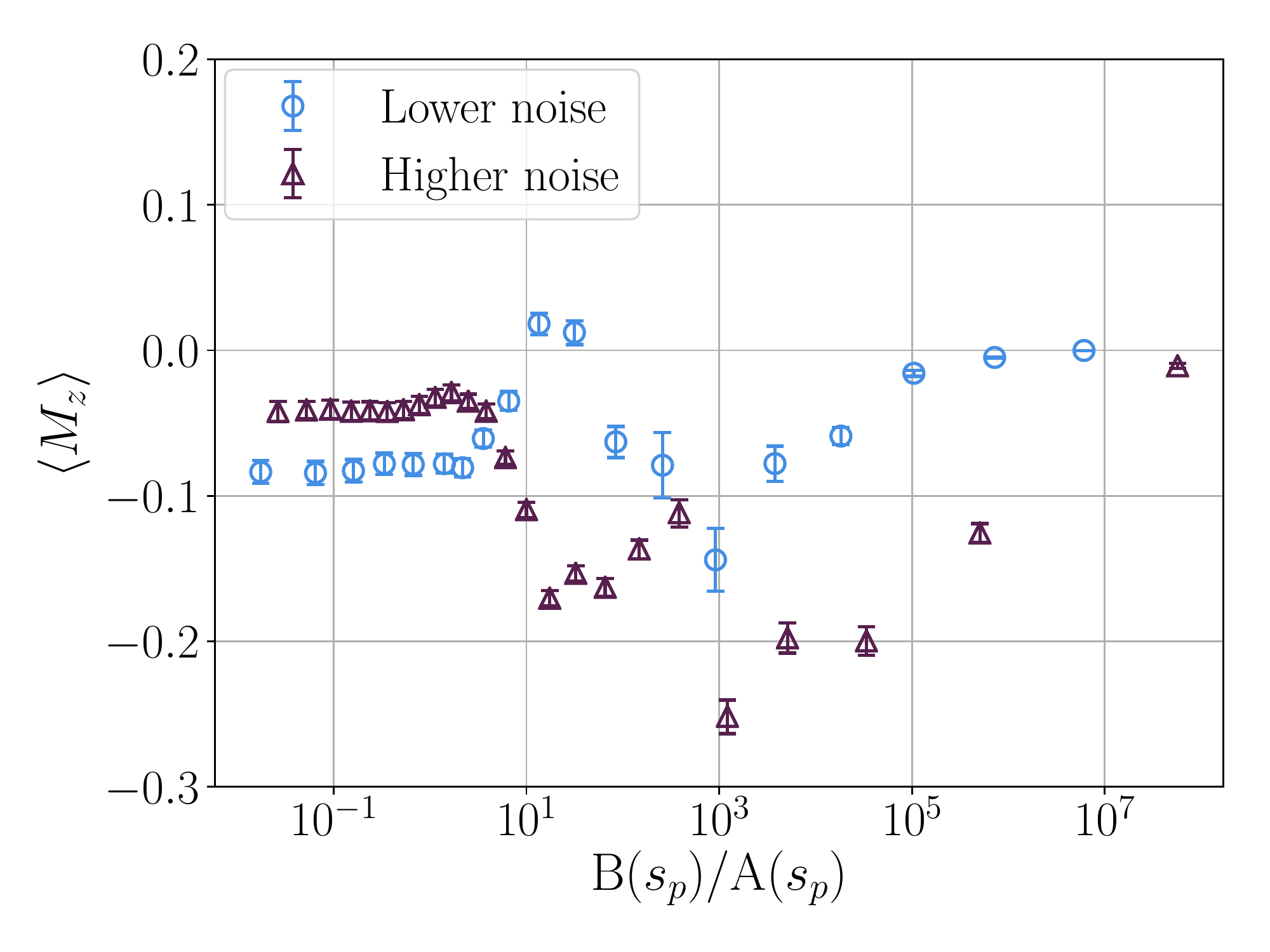}
\caption{Comparison between higher and lower noise processors: Total mean magnetization $\langle M_z \rangle$ for single-cell instances, for the initial states $[1, 1, 1, 1, -1, -1, -1, -1]$ (left) and $[1, 1, 1, -1, 1, -1, -1, -1]$ (right).}
\label{f14}
\end{figure}
\begin{figure}[!h]
\centering
\includegraphics[scale=0.34]{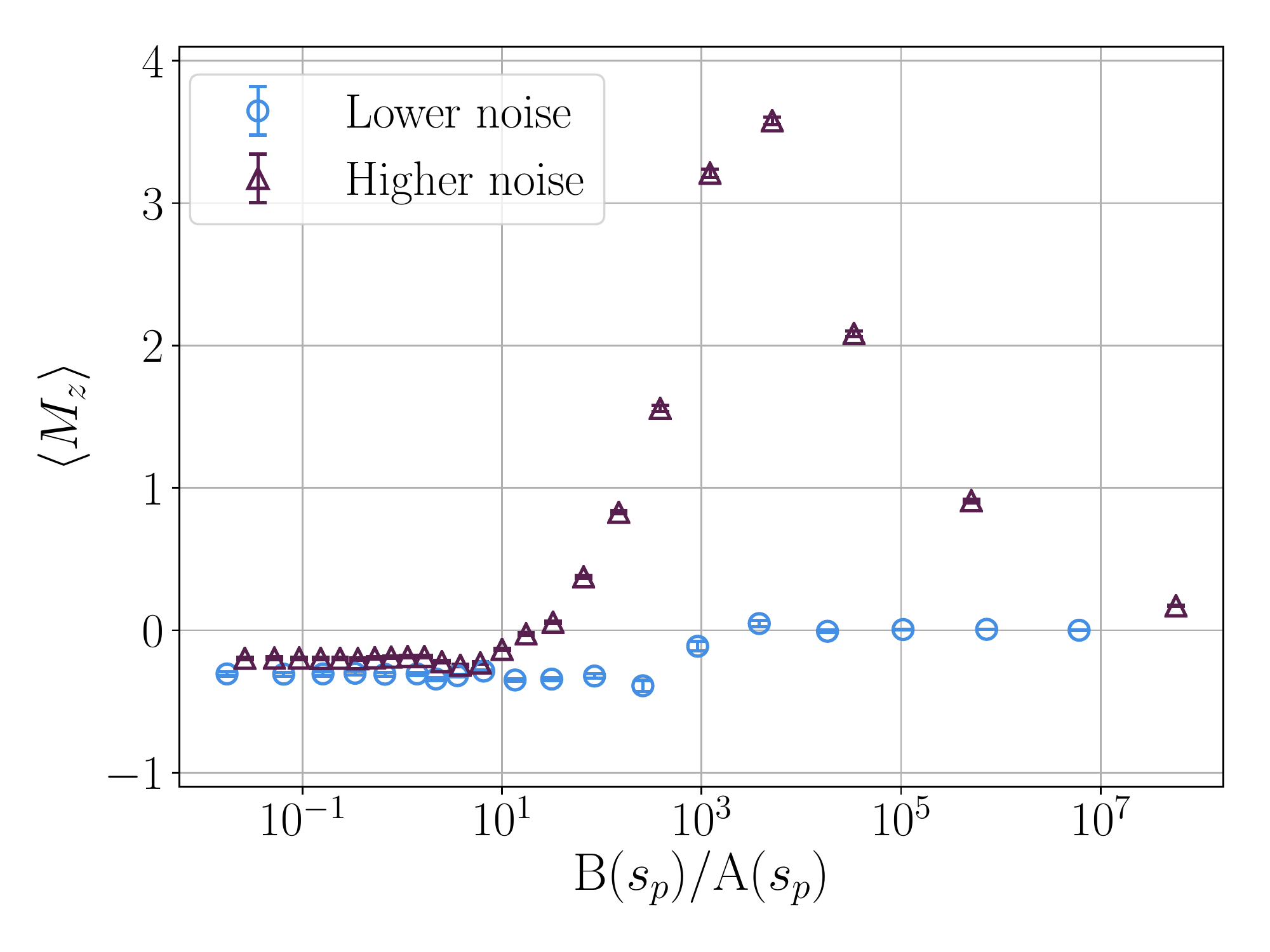}
\caption{Comparison between higher and lower noise processors: Total mean magnetization $\langle M_z \rangle$ for double-cell instances, for the initial state $[1, 1, 1, 1, -1, -1, -1, -1, 1, 1, 1, 1, -1, -1, -1, -1]$.}
\label{f15}
\end{figure}

\end{widetext}

A modification of SVMC has been proposed recently \cite{Albash:21} in order to better capture the slowing down of dynamics of the quantum annealer when the Ising Hamiltonian dominates over the transverse field \cite{Amin:15}.  When new angles are allowed to be picked in the range $[0, 2 \pi)$, the SVMC simulation can readily `jump' over large energy barriers that are present when $\mathrm{B}(s) \gg \mathrm{A}(s)$. To circumvent this problem, instead of proposing angles at random in the range $[0, 2 \pi)$, the new proposed angle is taken to depend on the annealing parameter $s$:
\begin{equation}
\theta_i' = \theta_i + \pi u \min(1.0,\mathrm{A}(s)/\mathrm{B}(s));
\end{equation}
where $u$ is a uniform random number in the range $[-1,1]$.  This update has the feature that when $\mathrm{A}(s) > \mathrm{B}(s)$ (when the transverse field is strong), the angles are updated randomly as in the original model.  When $\mathrm{A}(s) < \mathrm{B}(s)$ (when the Ising Hamiltonian dominates), the angle updates are localized more around the current value of the angle.  This has the feature that when $\mathrm{A}(s) \ll \mathrm{B}(s)$, the system is effectively frozen, which then qualitatively captures the expected `freeze-out' region for the quantum annealer \cite{Amin:15}. In principle, tuning how the range of proposed angles gets narrowed as $\mathrm{B}(s)/ \mathrm{A}(s)$ gets larger should allow us to get better quantitative agreement with the experimental results, but we do not pursue this here.

At the end of the anneal, if $\cos \theta_i > 0$, the rotor is projected onto the $1$ state, and if $\cos \theta_i < 0$, the rotor is projected onto the $-1$ state.

For our disordered simulations, we use 400 different noise realizations and 1000 independent SVMC simulations per noise realization.  This allows us to estimate for each noise realization the probability of each spin configuration and the expected magnetization.  Error bars of the magnetization in Figs. \ref{fig:SVMC} and \ref{fig:SVMC2} correspond to the 95\% confidence interval calculated using a bootstrap over the 400 expected magnetizations.

%%%%%%%%%%%%%%%%%%%% REFERENCES %%%%%%%%%%%%%%%%%%%%%
%\bibliography{biblioCORRE}


\end{document}